\documentclass[11pt,english]{article}
\usepackage[T1]{fontenc}
\usepackage[latin9]{inputenc}
\usepackage{color}
\usepackage{amsmath}
\usepackage{graphicx}
\usepackage{setspace}
\usepackage[authoryear]{natbib}

\makeatletter

\providecommand{\tabularnewline}{\\}

\@ifundefined{date}{}{\date{}}

\newtheorem{theorem}{Theorem}
\newtheorem{assumption}{Assumption}
\newtheorem{remark}{Remark}

\newcommand*{\indep}{%
\bot
}

\usepackage{babel}

\makeatother

\usepackage{babel}
\begin{document}

\title{Causal inference with observational studies trimmed by the estimated
propensity scores}

\author{Shu Yang\thanks{Department of Statistics, North Carolina State University, Raleigh,
North Carolina 27695, U.S.A.} and Peng Ding\thanks{Department of Statistics, University of California, Berkeley, California
94720, U.S.A.}}
\maketitle
\begin{abstract}
Causal inference with observational studies often relies on the assumptions
of unconfoundedness and overlap of covariate distributions in different
treatment groups. The overlap assumption is violated when some units
have propensity scores close to zero or one, and therefore both theoretical
and practical researchers suggest dropping units with extreme estimated
propensity scores. We advance the literature in three directions.
First, we clarify a conceptual issue of sample trimming by defining
causal parameters based on a target population without extreme propensity
score. Second, we propose a procedure of smooth weighting, which approximates
the existing sample trimming but has better asymptotic properties.
The new weighting estimator is asymptotically linear and the bootstrap
can be used to construct confidence intervals. Third, we extend the
theory to the average treatment effect on the treated, suggesting
trimming samples with estimated propensity scores close to one.

Some key words: Bootstrap; Lack of overlap; Non-smoothness; Potential
outcome; Unconfoundedness. 
\end{abstract}

\section{Introduction }

Under the potential outcomes framework \citep{rubin1974estimating},
causal effects are comparisons of the potential outcomes corresponding
to different treatments. There is an extensive literature on estimating
average treatment effects based on the assumption of unconfoundedness
and sufficient overlap in the covariate distributions \citep{rosenbaum1983central,imbens2015causal}.
Unfortunately, in many applications it is common to have limited overlap
in covariates between the treatment and control groups, i.e., there
are regions of the covariate space with low probability of receiving
treatment or control. Lack of overlap affects the credibility of all
methods attempting to estimate causal effects for the common population.
A consequence in weighting \citep{rosenbaum1983central,imbens2015causal}
is that extreme propensity scores induce substantively large weights,
which can result in a large variance and poor finite sample properties
\citep{kang2007demystifying,khan2010irregular}. In this case, it
is desirable to modify the estimand to averaging only over the part
of the covariate space with all treatment probabilities away from
zero. For example, \citet{crump2009dealing} suggested dropping subjects
from the analysis with estimated propensity score close to zero and
one, which generally alters the estimand by changing the reference
population \citep{li2016balancing}. In the current practice, researchers
often first trim the samples based on the estimated propensity scores,
and then characterize the target population and estimand based on
the sample estimates. This ad hoc definition of the treatment effect
is problematic, because using different samples may change the target
estimand.

The objective of this article is to clarify a conceptual issue of
sample trimming, which arises frequently in practice, leading to unambiguous
definitions of causal parameters based on a well-defined target population.
The non-smooth nature of trimming makes inference complicated. Therefore,
instead of making binary decisions to include or exclude subjects
from analysis, we propose to use a smooth weighting function so that
all subjects are weighted continuously. This smooth weighting approximates
the existing sample trimming, but allows us to derive the asymptotic
properties of the corresponding causal estimators using conventional
linearization methods for two-step statistics. We formally show that
the new weighting estimator is asymptotically linear and the bootstrap
can be used to construct confidence intervals. Moreover, by smoothing
the indicator function, the resulting estimators gain precision, as
demonstrated in the asymptotic analysis and simulation study. In
addition to the average treatment effect, we extend \citet{crump2009dealing}
to develop an optimal rule to select subpopulation for which the average
treatment effect on the treated can be estimated most precisely, and
establish asymptotic inference when support reduction is based estimated
propensity scores.\textcolor{blue}{} 

\section{Notation }

For each subject $i$, the treatment is $A_{i}\in\{0,1\}$, where
$0$ and $1$ are labels for control and treatment. There are two
potential outcomes, one for treatment and the other for control, denoted
by $Y_{i}(1)$ and $Y_{i}(0)$, respectively. The observed outcome
is $Y_{i}=Y_{i}(A_{i})$. Let $X_{i}$ be the observed pre-treatment
confounders. We assume that $\{A_{i},X_{i},Y_{i}(1),Y_{i}(0)\}_{i=1}^{N}$
are independent draws from the distribution of $\{A,X,Y(1),Y(0)\}$.
Given the observed confounders $X$, the conditional average causal
effect is $\tau(X)=E\{Y(1)-Y(0)\mid X\}$. The average treatment effect
is $\tau=E\{\tau(X)\}$, where the expectation is taken with respect
to the whole population. The common assumptions to identify $\tau$
are as follows \citep{rosenbaum1983central}.

\begin{assumption}[Unconfoundedness] $Y(a)\indep A\mid X$ for $a=0,1$.
\end{assumption}

\begin{assumption}[Sufficient overlap] There exist constants $c_{1}$
and $c_{2}$ such that with probability $1$, $0<c_{1}\leq e(X)\leq c_{2}<1$,
where $e(X)=\mathrm{pr}(A=1\mid X)$ is the propensity score. \end{assumption}

\begin{assumption}$E\{Y(a)^{2}\}<\infty$, for $a=0,1$. \end{assumption}

In observational studies the propensity score is not known and therefore
has to be estimated from data. Following \citet{rosenbaum1983central}
and most of the empirical literature, we assume that the propensity
score is correctly specified by a generalized linear model $e(X)=e(X'\theta^{*})$.
Let $\hat{\theta}$ be the maximum likelihood estimator of $\theta^{*}$.
Our method is also applicable to other asymptotically linear estimators
of $\theta$. Then, a simple weighting estimator of $\tau$ is $N^{-1}\sum_{i=1}^{N}\hat{\tau}(X_{i})$,
where
\begin{equation}
\hat{\tau}(X_{i})=\frac{A_{i}Y_{i}}{e(X_{i}'\hat{\theta})}-\frac{(1-A_{i})Y_{i}}{1-e(X_{i}'\hat{\theta})}.\label{eq:weightE1}
\end{equation}
The augmented weighting estimator \citep{lunceford2004stratification,bang2005doubly}
augments the simple weighting estimator by further estimating $\mu(a,X)=E(Y\mid A=a,X)$
by $\hat{\mu}(a,X)$, and using $N^{-1}\sum_{i=1}^{N}\hat{\tau}^{\mathrm{aug}}(X_{i}),$
where 
\begin{multline}
\hat{\tau}^{\mathrm{aug}}(X_{i})=\left[\frac{A_{i}Y_{i}}{e(X_{i}'\hat{\theta})}+\left\{ 1-\frac{A_{i}}{e(X_{i}'\hat{\theta})}\right\} \hat{\mu}(1,X_{i})\right]\\
-\left[\frac{(1-A_{i})Y_{i}}{1-e(X_{i}'\hat{\theta})}+\left\{ 1-\frac{1-A_{i}}{1-e(X_{i}'\hat{\theta})}\right\} \hat{\mu}(0,X_{i})\right].\label{eq:weightE2}
\end{multline}
The augmented weighting estimator features a double robustness property
in the sense that under Assumptions 1\textendash 3, if either $e(X)$
or $\mu(a,X)$ is correctly specified, $\hat{\tau}^{\mathrm{aug}}$
is consistent for $\tau$. 

The weighting estimators suffer from large variability especially
when Assumption 2 is violated or close to be violated. In the presence
of lack of overlap, define the set with sufficient overlap to be $\mathcal{O}=\{x\mid\alpha\leq e(x)\leq1-\alpha\}$,
where $\alpha$ is a fixed cut-off value, e.g., a rule of thumb is
$\alpha=0.1$, as suggested by \citet{crump2009dealing}. The target
population is then represented by $\mathcal{O}$, and the estimand
of interest becomes $\tau(\mathcal{O})=E\{\tau(X)\mid X\in\mathcal{O}\}$.
This estimand does not depend on the sample, which is more straightforward
to interpret. 

In existing sampling trimming, the inclusion weight is
\begin{equation}
\omega(X_{i}'\hat{\theta})=1_{\{X_{i}\in\hat{\mathcal{O}}\}}=1_{\{\alpha\leq e(X_{i}'\hat{\theta})\leq1-\alpha\}},\label{eq:indicator weight}
\end{equation}
where $1_{\{\cdot\}}$ being the indicator function, and $\hat{\mathcal{O}}=\{x\mid\alpha\leq e(x'\hat{\theta})\leq1-\alpha\}$
is the trimmed sample based on the estimated propensity scores. The
weighting estimators of $\tau(\mathcal{O})$ become 
\begin{equation}
\hat{\tau}(\hat{\theta})=\left\{ \sum_{i=1}^{N}\omega(X_{i}'\hat{\theta})\right\} ^{-1}\sum_{i=1}^{N}\omega(X_{i}'\hat{\theta})\hat{\tau}(X_{i}),\label{eq:estimator1}
\end{equation}
\begin{equation}
\hat{\tau}^{\mathrm{aug}}(\hat{\theta})=\left\{ \sum_{i=1}^{N}\omega(X_{i}'\hat{\theta})\right\} ^{-1}\sum_{i=1}^{N}\omega(X_{i}'\hat{\theta})\hat{\tau}^{\mathrm{aug}}(X_{i}),\label{eq:estimator2}
\end{equation}
where $\hat{\tau}(X_{i})$ and $\hat{\tau}^{\mathrm{aug}}(X_{i})$
are defined in (\ref{eq:weightE1}) and (\ref{eq:weightE2}), respectively.
We write $\hat{\tau}=\hat{\tau}(\hat{\theta})$ and $\hat{\tau}^{\mathrm{aug}}=\hat{\tau}^{\mathrm{aug}}(\hat{\theta})$
in shorthand. 

The main question addressed in this article is how the estimated support
affects the inference. To study the asymptotic behaviors of $\hat{\tau}$
and $\hat{\tau}^{\mathrm{aug}}$, we need to take into account of
first the sampling variability in $\hat{\theta}$, which induces variability
of the estimated set $\hat{\mathcal{O}}$ and second the sampling
variability in $\hat{\tau}$ and $\hat{\tau}^{\mathrm{aug}}$. We
can not directly apply conventional asymptotic linearization methods
because the weight function (\ref{eq:indicator weight}) is non-smooth.
To avoid this difficulty, we consider a smoothed version of the weight
function 
\begin{equation}
\omega_{\epsilon}(X_{i}'\hat{\theta})=\Phi_{\epsilon}\left\{ e(X_{i}'\hat{\theta})-\alpha\right\} \Phi_{\epsilon}\left\{ 1-\alpha-e(X_{i}'\hat{\theta})\right\} ,\label{eq:smooth weight}
\end{equation}
where $\Phi_{\epsilon}(z)$ is a normal cumulative distribution with
mean zero and variance $\epsilon$. The new weight function imposes
a soft threshold, instead of a hard threshold which results in weights
either zero or one, for the estimated propensity scores close to $\alpha$
and $1-\alpha$. An important issue regarding this weight function
is the choice of $\epsilon$. As $\epsilon\rightarrow0$, the smooth
weight function (\ref{eq:smooth weight}) coverages to the indicator
weight function (\ref{eq:indicator weight}); see Figure S1 in the
Supplementary Material for visualization of the weight functions.
Therefore, for small $\epsilon$, the behaviors of the estimators
(\ref{eq:estimator1}) and (\ref{eq:estimator2}) with the smooth
weight function (\ref{eq:smooth weight}) are similar to those with
the indicator weight function (\ref{eq:indicator weight}). We derive
the asymptotic results for the smoothed estimators. 

\section{Main Results}

Based on data $\{(A_{i},X_{i})\}_{i=1}^{N}$, let the score function
and the Fisher information matrix of $\theta$ be 
\[
S(\theta)=\frac{1}{N}\sum_{i=1}^{N}X_{i}\frac{A_{i}-e(X_{i}'\theta)}{e(X_{i}'\theta)\{1-e(X_{i}'\theta)\}}f(X_{i}'\theta),\quad\mathcal{I}_{\theta}=E\left[\frac{f(X'\theta)^{2}}{e(X'\theta)\{1-e(X'\theta)\}}XX'\right],
\]
respectively, where $f(t)=\mathrm{d}e(t)/\mathrm{d}t$. Because $\hat{\theta}$
is the solution to the score equation $S(\theta)=0$, under certain
regularity conditions, $\hat{\theta}-\theta^{*}=\mathcal{I}_{\theta^{*}}^{-1}S(\theta^{*})+o_{p}(N^{-1/2}).$
Let $\sigma^{2}(a,X)=\mathrm{var}(Y\mid A=a,X)$, for $a=0,1$. Denote
$\hat{\tau}_{\epsilon}$ to be the weighting estimator (\ref{eq:estimator1})
with the smooth weight function (\ref{eq:smooth weight}), and $\tau_{\epsilon}=E\{\omega_{\epsilon}(X'\theta^{*})\tau(X)\}$. 

\begin{theorem}\label{main1} Under Assumptions 1 and 3, $\hat{\tau}_{\epsilon}$
is asymptotically linear. Moreover, 
\[
N^{1/2}(\hat{\tau}_{\epsilon}-\tau_{\epsilon})\rightarrow\mathcal{N}\left(0,\sigma_{\epsilon}^{2}+b_{1,\epsilon}'\mathcal{I}_{\theta^{*}}^{-1}b_{1,\epsilon}-b_{2,\epsilon}'\mathcal{I}_{\theta^{*}}^{-1}b_{2,\epsilon}\right),
\]
in distribution, as $N\rightarrow\infty$, where
\begin{equation}
b_{1,\epsilon}=E\left\{ \frac{\partial}{\partial\theta}\left[\frac{\omega_{\epsilon}(X'\theta^{*})}{E\{\omega_{\epsilon}(X'\theta^{*})\}}\right]\tau(X)\right\} ,\label{eq:b1}
\end{equation}
\[
b_{2,\epsilon}=\frac{1}{E\{\omega_{\epsilon}(X'\theta^{*})\}}E\left\{ \omega_{\epsilon}(X'\theta^{*})f(X'\theta^{*})\left[\frac{E\{X\mu(1,X)\mid e(X)\}}{e(X)}+\frac{E\{X\mu(0,X)\mid e(X)\}}{1-e(X)}\right]\right\} ,
\]
\begin{eqnarray*}
\sigma_{\epsilon}^{2} & = & 
\frac{1}{E\{\omega_{\epsilon}(X'\theta^{*})\}^{2}}	\mathrm{var}\left\{   \omega_\epsilon(X' \theta^*)\tau(X)   \right\}\\
 &  & +\frac{1}{E\{\omega_{\epsilon}(X'\theta^{*})\}^{2}}E\left\{ \omega_{\epsilon}(X'\theta^{*})^{2}\left[\left\{ \frac{1-e(X)}{e(X)}\right\} ^{1/2}\mu(1,X)+\left\{ \frac{e(X)}{1-e(X)}\right\} ^{1/2}\mu(0,X)\right]^{2}\right\} \\
 &  & +\frac{1}{E\{\omega_{\epsilon}(X'\theta^{*})\}^{2}}E\left[\omega_{\epsilon}(X'\theta^{*})^{2}\left\{ \frac{\sigma^{2}(1,X)}{e(X)}+\frac{\sigma^{2}(0,X)}{1-e(X)}\right\} \right].
\end{eqnarray*}
\end{theorem}

\begin{remark}\label{rmk1}

The term $b_{1,\epsilon}'\mathcal{I}_{\theta^{*}}^{-1}b_{1,\epsilon}$
is the increased variability due to estimating the support. We now
show that this term is close to zero for small $\epsilon$. We note
\begin{equation}
\frac{\partial}{\partial\theta}\left[\frac{\omega_{\epsilon}(X'\theta^{*})}{E\{\omega_{\epsilon}(X'\theta^{*})\}}\right]=\frac{\dot{\omega}_{\epsilon}(X'\theta^{*})E\{\omega_{\epsilon}(X'\theta^{*})\}-E\{\dot{\omega}_{\epsilon}(X'\theta^{*})\}\omega_{\epsilon}(X'\theta^{*})}{[E\{\omega_{\epsilon}(X'\theta^{*})\}]^{2}},\label{eq:b1-1}
\end{equation}
where
\begin{eqnarray*}
\dot{\omega}_{\epsilon}(X'\theta^{*}) & = & \frac{\partial}{\partial\theta}\left[\Phi_{\epsilon}\left\{ e(X'\theta^{*})-\alpha\right\} \Phi_{\epsilon}\left\{ 1-\alpha-e(X'\theta^{*})\right\} \right]\\
 & = & \phi_{\epsilon}\left\{ e(X'\theta^{*})-\alpha\right\} \Phi_{\epsilon}\left\{ 1-\alpha-e(X'\theta^{*})\right\} f(X'\theta^{*})X\\
 &  & -\Phi_{\epsilon}\left\{ e(X'\theta^{*})-\alpha\right\} \phi_{\epsilon}\left\{ 1-\alpha-e(X'\theta^{*})\right\} f(X'\theta^{*})X,
\end{eqnarray*}
and $\phi_{\epsilon}(x)=\mathrm{d}\Phi_{\epsilon}(x)/\mathrm{d}x$.
As $\epsilon\rightarrow0$, because $\phi_{\epsilon}(x)\rightarrow0$,
the right hand side of (\ref{eq:b1-1}), and therefore $b_{1,\epsilon}$,
go to zero. The increased variability due to the estimated support
is close to zero with small $\epsilon$.

\end{remark}

\begin{remark} The term $-b_{2,\epsilon}'\mathcal{I}_{\theta}^{-1}b_{2,\epsilon}$
implies that the estimated propensity score increases the precision
of the simple weighting estimator of $\tau$ based on the true propensity
score, which has been demonstrated in the missing data and causal
literature; see, e.g., \citet{rubin1992affinely} and \citet{abadie2016matching}.
\end{remark}

\begin{remark}

Assuming that $\tau(X)$ is integrable, by the Dominated Convergence
Theorem, $\tau_{\epsilon}$ converges to $\tau(\mathcal{O})$ as $\epsilon\rightarrow0$.
This implies that our inference based on $\hat{\tau}_{\epsilon}$,
by choosing a small $\epsilon$, can be drawn for the target population
represented by $\mathcal{O}$.

\end{remark}

Let $\hat{\tau}_{\epsilon}^{\mathrm{aug}}$ be the weighting estimator
(\ref{eq:estimator2}) with the smooth weight function (\ref{eq:smooth weight}).

\begin{theorem}\label{main2}Under Assumptions 1 and 3, $\hat{\tau}_{\epsilon}^{\mathrm{aug}}$
is asymptotically linear. Moreover, 
\[
N^{1/2}(\hat{\tau}_{\epsilon}^{\mathrm{aug}}-\tau_{\epsilon})\rightarrow\mathcal{N}\left\{ 0,\tilde{\sigma}_{\epsilon}^{2}+b_{1\epsilon}'\mathcal{I}_{\theta^{*}}b_{1\epsilon}+(C_{0}+C_{1})'\mathcal{I}_{\theta^{*}}(C_{0}+C_{1})+\tilde{B}'(C_{0}-C_{1})\right\} ,
\]
in distribution, as $N\rightarrow\infty$, where $b_{1,\epsilon}$
is defined in (\ref{eq:b1}),\textbf{\textcolor{black}{{} 
\begin{eqnarray*}
\tilde{\sigma}_{\epsilon}^{2} & = & 
\frac{1}{E\{\omega_{\epsilon}(X'\theta^{*})\}^{2}}	\mathrm{var}\left\{   \omega_\epsilon(X' \theta^*)\tau(X)   \right\}\\
 &  & +\frac{1}{E\{\omega_{\epsilon}(X'\theta^{*})\}^{2}}E\left[\omega_{\epsilon}(X'\theta^{*})^{2}\left\{ \frac{\sigma^{2}(1,X)}{e(X)}+\frac{\sigma^{2}(0,X)}{1-e(X)}\right\} \right],
\end{eqnarray*}
}}
\[
C_{0}=E\left\{ X\omega_{\epsilon}(X'\theta^{*})f(X'\theta^{*})\frac{\tilde{\mu}(0,X)-\mu(0,X)}{1-e(X)}\right\} ,
\]
\[
C_{1}=E\left\{ X\omega_{\epsilon}(X'\theta^{*})f(X'\theta^{*})\frac{\tilde{\mu}(1,X)-\mu(1,X)}{e(X)}\right\} ,
\]
with $\hat{\mu}(a,X)\rightarrow\tilde{\mu}(a,X)$ in probability,
for $a=0,1$, and $\tilde{B}=b_{1,\epsilon}-C_{0}-C_{1}$.

\end{theorem}

\begin{remark}\label{rmk2} The term $b_{1,\epsilon}'\mathcal{I}_{\theta^{*}}^{-1}b_{1,\epsilon}$
can be made small by choosing a small $\epsilon$, as shown in Remark
\ref{rmk1}. If the outcome model is correctly specified, $\tilde{\mu}(a,X)=\mu(a,X)$,
and consequently, $C_{0}=C_{1}=0$. The asymptotic variance of $\hat{\tau}_{\epsilon}^{\mathrm{aug}}$
reduces to $\tilde{\sigma}_{\epsilon}^{2}+b_{1\epsilon}'\mathcal{I}_{\theta^{*}}b_{1\epsilon}$\textcolor{black}{,
which is more efficient than $\hat{\tau}_{\epsilon}$. Intuitively,
this occurs because by regressing $Y$ on $X$ and $A$, we are essentially
using the residual as the new outcome, which in general has smaller
variance than $Y$. }

\end{remark}

\begin{remark}Because $\hat{\tau}_{\epsilon}$ and $\hat{\tau}_{\epsilon}^{\mathrm{aug}}$
are asymptotically linear, the bootstrap can be used to estimate the
variances of $\hat{\tau}_{\epsilon}$ and $\hat{\tau}_{\epsilon}^{\mathrm{aug}}$
\citep{shao2012jackknife}. Let $\mathcal{S}=\{x\mid e(x'\theta^{*})=\alpha$
or $1-\alpha\}$. If $\mathrm{pr}(X\in\mathcal{S})=0$, we conjecture
that the bootstrap works for the weighting estimator with the indicator
function. This is demonstrated in the simulation study. 

\end{remark}

\section{Average treatment effect on the treated}

Another estimand of interest is the average treatment effect for the
treated $\tau_{\mathrm{ATT}}=E\{Y(1)-Y(0)\mid A=1\}=E\{\tau(X)\mid A=1\}$
\citep{rubin1977assignment,hirano2001estimation}. The outcome distribution
for the treated is empirically identifiable $E\{Y(1)\mid A=1\}=E(Y\mid A=1)$,
and therefore Assumptions 1 and 2 can be weakened \citep{heckman1997matching}.

\begin{assumption}$Y(0)\indep A\mid X$. \end{assumption}

\begin{assumption}\label{assump2}There exists a constant $c_{2}$
such that with probability $1$, $e(X)\leq c_{2}<1$. \end{assumption}

A simple weighting estimator of $\tau_{\mathrm{ATT}}$ \citep{hirano2003efficient}
is 
\begin{equation}
\hat{\tau}_{\mathrm{ATT}}=\frac{\sum_{i=1}^{N}A_{i}Y_{i}}{\sum_{i=1}^{N}e(X_{i}'\hat{\theta})}-\frac{\sum_{i=1}^{N}(1-A_{i})Y_{i}e(X_{i}'\hat{\theta})/\{1-e(X_{i}'\hat{\theta})\}}{\sum_{i=1}^{N}e(X_{i}'\hat{\theta})}=\frac{\sum_{i=1}^{N}e(X_{i}'\hat{\theta})\hat{\tau}(X_{i})}{\sum_{i=1}^{N}e(X_{i}'\hat{\theta})}.\label{eq:ATT-weightE1}
\end{equation}
By the above expression, $\hat{\tau}_{\mathrm{ATT}}$ is a special
case of the weighting estimator (\ref{eq:estimator1}) by choosing
$\omega(X_{i}'\hat{\theta})=e(X_{i}'\hat{\theta})$. Analogously,
we propose the augmented weighting estimator of $\tau_{\mathrm{ATT}}$,
\begin{eqnarray}
\hat{\tau}_{\mathrm{ATT}}^{\mathrm{aug}} & = & \frac{\sum_{i=1}^{N}e(X_{i}'\hat{\theta})\hat{\tau}^{\mathrm{aug}}(X_{i})}{\sum_{i=1}^{N}e(X_{i}'\hat{\theta})}.\label{eq:ATT-weightE2}
\end{eqnarray}

There is a limited literature dealing with the lack of overlap for
$\tau_{\mathrm{ATT}}$ when Assumption \ref{assump2} may not hold.
Similar to \citet{crump2009dealing}, assuming that $\sigma^{2}(1,X)=\sigma^{2}(0,X)$,
we can show that the optimal overlap for estimating $\tau_{\mathrm{ATT}}$
is of the form $\mathcal{O}=\{x\mid1-e(x)\geq\alpha\}$ for some $\alpha$,
for which the estimators have smallest asymptotic variance. Intuitively,
for the treated subjects with $e(X)$ close to one, there are no similar
subjects in the control group that can provide adequate information
to infer $Y(0)$ for these treated subjects. Statistically, the control
subjects with $e(X)$ close to one contribute to large weights. Therefore,
it is reasonable to drop these subjects with $e(X)$ close to one.
By restricting the focus to the optimal set, the estimand of interest
becomes $\tau_{\mathrm{ATT}}(\mathcal{O})=E\{\tau(X)\mid A=1,X\in\mathcal{O}\}$,
for which we propose two estimators with smooth inclusion weights
as
\begin{eqnarray*}
\hat{\tau}_{\mathrm{ATT},\epsilon} & = & \frac{\sum_{i=1}^{N}\Phi_{\epsilon}\{1-\alpha-e(X_{i}'\hat{\theta})\}e(X_{i}'\hat{\theta})\hat{\tau}(X_{i})}{\sum_{i=1}^{N}\Phi_{\epsilon}\{1-\alpha-e(X_{i}'\hat{\theta})\}e(X_{i}'\hat{\theta})},\\
\hat{\tau}_{\mathrm{ATT},\epsilon}^{\mathrm{aug}} & = & \frac{\sum_{i=1}^{N}\Phi_{\epsilon}\{1-\alpha-e(X_{i}'\hat{\theta})\}e(X_{i}'\hat{\theta})\hat{\tau}^{\mathrm{aug}}(X_{i})}{\sum_{i=1}^{N}\Phi_{\epsilon}\{1-\alpha-e(X_{i}'\hat{\theta})\}e(X_{i}'\hat{\theta})}.
\end{eqnarray*}
The asymptotic properties can be derived similarly as in Theorems
1 and 2 by recognizing $\omega_{\epsilon}(X'\hat{\theta})=\Phi_{\epsilon}\{1-\alpha-e(X_{i}'\hat{\theta})\}e(X_{i}'\hat{\theta})$
for $\hat{\tau}_{\mathrm{ATT},\epsilon}$ and $\hat{\tau}_{\mathrm{ATT},\epsilon}^{\mathrm{aug}}$.
In particular, the asymptotic linearity enables the bootstrap for
inference. See the Supplementary Material for details. Similar discussion
applies to estimating the average treatment effect on the control. 

\section{Simulation study}

We assess the performance of the new weighting estimators of the average
treatment effect over a target population. We consider six covariates
$X_{j}$ $(j=1,\ldots,6)$, where $X_{1}$, $X_{2}$, and $X_{3}$
are multivariate normal with means $(0,0,0)$, variances $(2,1,1)$
and covariances $(1,-1,-0.5)$, $X_{4}\sim\mathrm{Uniform}[-3,3]$,
$X_{5}\sim\chi_{1}^{2}$, and $X_{6}\sim\mathrm{Bernoulli}(0.5)$.
Let $X=(X_{1},X_{2},X_{3},X_{4},X_{5},X_{6})'$ be the 6-component
vector of covariates. The treatment indicator $A$ is generated from
a Bernoulli distribution with probability $e(X)$. We consider four
propensity score deigns: 
\begin{description}
\item [{(P1)}] $e(X)=\mathrm{logit}\{0.1\times(X_{1}+X_{2}+X_{3}+X_{4}+X_{5}+X_{6})\}$. 
\item [{(P2)}] $e(X)=\mathrm{logit}\{0.8\times(X_{1}+X_{2}+X_{3}+X_{4}+X_{5}+X_{6})\}$.
\item [{(P3)}] $e(X)=\mathrm{logit}\{0.1\times(X_{1}+X_{2}^{2}+X_{3}^{2}+X_{4}+X_{5}+X_{6})\}$.
\item [{(P4)}] $e(X)=\mathrm{logit}\{0.8\times(X_{1}+X_{2}^{2}+X_{3}^{2}+X_{4}+X_{5}+X_{6})\}$.
\end{description}
(P1) and (P3) represent week separations of propensity score distributions
between the treatment and control groups; (P2) and (P4) represent
strong separations. See Figure S.2 in the Supplementary Material for
visualization of propensity score distributions. We consider two outcome
designs: 
\begin{description}
\item [{(O1)}] $Y(a)=a(X_{1}+X_{2}+X_{3}-X_{4}+X_{5}+X_{6})+\eta$, with
$\eta\sim\mathcal{N}(0,1)$, for $a=0,1$. 
\item [{(O2)}] $Y(a)=a(X_{1}+X_{2}+X_{3})^{2}+\eta$, with $\eta\sim\mathcal{N}(0,1)$,
for $a=0,1$. 
\end{description}
(O1) is a linear case, and (O2) is a non-linear case. The sample size
is set to be $N=500$. The target population is represented by $\mathcal{O}=\{x\mid\alpha<e(x)<1-\alpha\}$
with $\alpha=0.1$, and the estimand of interest $\tau(\mathcal{O})$
is the average treatment effect over the target population.

We consider the weighting estimators with indicator weight function
and smooth weight function, and $\tau(\mathcal{O})=(\sum_{i=1}^{N}1_{\{X_{i}\in\mathcal{O}\}})^{-1}\sum_{i=1}^{N}1_{\{X_{i}\in\mathcal{O}\}}\{Y_{i}(1)-Y_{i}(0)\}$
for benchmark comparison. The propensity scores are estimated by a
logistic regression model with linear predictors $X$. Therefore,
the propensity score model is correctly specified under (P1) and (P2),
but it is misspecified under (P3) and (P4). For the augmented weighting
estimators, $\mu(a,X)$ is estimated by a simple regression of $Y$
on $X$, separately for $A=a$ ($a=0,1$). Therefore, the outcome
regression model is correctly specified under (O1) but misspecified
under (O2).

\begin{table}
\global\long\def\arraystretch{1.3}
 \protect\caption{\label{tab:Results-1}Results: mean, variance (var), and variance
estimate (ve) by $100$ bootstrapping under eight combinations of
outcome design and propensity score design: c indicates the corresponding
model is correctly specified , and w indicates the corresponding model
is incorrectly specified }
\centering{}
 \resizebox{\columnwidth}{!}{ 
\begin{tabular}{cccccccccccccc}
{Scenario}  &  & \multicolumn{3}{c}{{i (OD1c, PSD1c)}} & \multicolumn{3}{c}{{ii (OD1c, PSD2c)}} & \multicolumn{3}{c}{{iii (OD1c, PSD3w) }} & \multicolumn{3}{c}{{iv (OD1c, PSD4w)}}\tabularnewline
 & {$\epsilon$}  & {mean}  & {var}  & {ve}  & {mean}  & {var}  & {ve}  & {mean}  & {var}  & {ve}  & {mean}  & {var}  & {ve}\tabularnewline
{$\tau(\mathcal{O})$}  &  & {1.46}  &  &  & {1.33 }  &  &  & {1.44 }  &  &  & {1.37}  &  & \tabularnewline
{$\hat{\tau}(\hat{\theta})$}  & {\textendash }  & {1.45}  & \textcolor{black}{0.0341}  & {0.0336 }  & {1.33}  & \textcolor{black}{0.0471}  & {0.0518}  & {1.48 }  & \textcolor{black}{0.0285}  & {0.0282}  & {1.45}  & \textcolor{black}{0.0399}  & {0.0405}\tabularnewline
{$\hat{\tau}^{\mathrm{aug}}(\hat{\theta})$}  & {\textendash }  & {1.46}  & \textcolor{black}{0.0282}  & {0.0267}  & {1.32}  & \textcolor{black}{0.0343}  & {0.0342}  & {1.50 }  & \textcolor{black}{0.0263}  & {0.0253}  & {1.49}  & \textcolor{black}{0.0331}  & {0.0315}\tabularnewline
{$\hat{\tau}_{\epsilon}(\hat{\theta})$}  & {$10^{-4}$}  & {1.45}  & \textcolor{black}{0.0333}  & {0.0331}  & {1.33}  & \textcolor{black}{0.0445}  & {0.0474}  & {1.48 }  & \textcolor{black}{0.0284}  & {0.0278}  & {1.45}  & \textcolor{black}{0.0386}  & {0.0382}\tabularnewline
{$\hat{\tau}_{\epsilon}^{\mathrm{aug}}(\hat{\theta})$}  & {$10^{-4}$}  & {1.46}  & \textcolor{black}{0.0280}  & {0.0267}  & {1.33}  & \textcolor{black}{0.0339}  & {0.0333}  & {1.50 }  & \textcolor{black}{0.0263}  & {0.0252}  & {1.49}  & \textcolor{black}{0.0327}  & {0.0308}\tabularnewline
{$\hat{\tau}_{\epsilon}(\hat{\theta})$}  & {$10^{-5}$}  & {1.45}  & \textcolor{black}{0.0339}  & {0.0331}  & {1.33}  & \textcolor{black}{0.0464}  & {0.0503}  & {1.48 }  & \textcolor{black}{0.0285 }  & {0.0281}  & {1.45}  & \textcolor{black}{0.0394}  & {0.0397}\tabularnewline
{$\hat{\tau}^{\mathrm{aug}}(\hat{\theta})$}  & {$10^{-5}$}  & {1.46}  & \textcolor{black}{0.0282}  & {0.0267}  & {1.32}  & \textcolor{black}{0.0343}  & {0.0342}  & {1.50 }  & \textcolor{black}{0.0263}  & {0.0253}  & {1.49}  & \textcolor{black}{0.0331}  & {0.0315}\tabularnewline
 &  & \multicolumn{3}{c}{{v (OD2w, PSD1c)}} & \multicolumn{3}{c}{{vi (OD2w, PSD2c)}} & \multicolumn{3}{c}{{vii (OD2w, PSD3w) }} & \multicolumn{3}{c}{{viii (OD2w, PSD4w)}}\tabularnewline
{$\tau(\mathcal{O})$}  &  & {7.58 }  &  &  & {6.69}  &  &  & {7.62 }  &  &  & {5.96 }  &  & \tabularnewline
{$\hat{\tau}(\hat{\theta})$}  & {\textendash }  & {7.58 }  & \textcolor{black}{0.9400}  & {0.8912}  & {6.69}  & {0.8983}  & {0.9811}  & \textcolor{black}{8.75}  & {0.9201}  & {0.9122}  & \textcolor{black}{8.93}  & {1.4198}  & {1.3808}\tabularnewline
{$\hat{\tau}^{\mathrm{aug}}(\hat{\theta})$}  & {\textendash }  & {7.59 }  & \textcolor{black}{0.8538}  & {0.7652}  & {6.67}  & {0.7919}  & {0.8417}  & \textcolor{black}{8.82}  & {0.8493}  & {0.7925}  & \textcolor{black}{9.06}  & {1.2260}  & {1.0958 }\tabularnewline
{$\hat{\tau}_{\epsilon}(\hat{\theta})$}  & {$10^{-4}$}  & {7.57 }  & \textcolor{black}{0.8861}  & {0.8408}  & {6.70}  & {0.8528 }  & {0.8967 }  & \textcolor{black}{8.75}  & {0.9106}  & {0.8828}  & \textcolor{black}{8.94}  & {1.3418}  & {1.2842}\tabularnewline
{$\hat{\tau}_{\epsilon}^{\mathrm{aug}}(\hat{\theta})$}  & {$10^{-4}$}  & {7.58}  & \textcolor{black}{0.8268}  & {0.7473}  & {6.68}  & {0.7663 }  & {0.7941}  & \textcolor{black}{8.82}  & {0.8441}  & {0.7839}  & \textcolor{black}{9.07}  & {1.1896}  & {1.0554}\tabularnewline
{$\hat{\tau}_{\epsilon}(\hat{\theta})$}  & {$10^{-5}$}  & {7.57 }  & \textcolor{black}{0.9203}  & {0.8732}  & {6.69}  & {0.8879 }  & {0.9525 }  & \textcolor{black}{8.75}  & {0.9192}  & {0.9020}  & \textcolor{black}{8.93}  & {1.3997}  & {1.3479}\tabularnewline
{$\hat{\tau}_{\epsilon}^{\mathrm{aug}}(\hat{\theta})$}  & {$10^{-5}$}  & {7.59}  & \textcolor{black}{0.8405 }  & {0.7591 }  & {6.68}  & {0.7868 }  & {0.8249}  & \textcolor{black}{8.82}  & {0.8474}  & {0.7896}  & \textcolor{black}{9.06}  & {1.2171}  & {1.0824}\tabularnewline
\end{tabular}} 
 
\end{table}

Table \ref{tab:Results-1} shows the simulation results. Under Scenarios
i, ii, v and vi when the propensity score model is correctly specified,
the weighting estimators are unbiased of $\tau(\mathcal{O})$, and
the augmented weighting estimators therefore improve the precision.
However, under Scenarios iii, iv, vii and viii when the propensity
score model is misspecified, all estimators are biased even when the
outcome regression model is correctly specified for the augmented
weighting estimators. The augmented weighting estimators are not doubly
robust in this case, because selecting samples corresponding to the
target population relies on correct specification of the propensity
score model. We further address the misspecification of propensity
score model in the discussion section. The weighting estimators with
smooth inclusion weights, $\hat{\tau}_{\epsilon}$ and $\hat{\tau}_{\epsilon}^{\mathrm{aug}}$,
show slightly smaller variances than the counterparts with indicator
inclusion weights, $\hat{\tau}$ and $\hat{\tau}^{\mathrm{aug}}$.
Moreover, as $\epsilon$ becomes smaller, the performances of $\hat{\tau}_{\epsilon}$
and $\hat{\tau}_{\epsilon}^{\mathrm{aug}}$ become closer to those
of $\hat{\tau}$ and $\hat{\tau}^{\mathrm{aug}}$. The bootstrap works
well with variance estimates close to the true variances for all estimators
including the weighting estimators with indicator inclusion weights.

In addition, we illustrate our method using two real-life data sets,
presented in the Supplementary Material. 

\section{Discussion}

The propensity score model is critical for our weighting estimators.
The majority of the literature used a parametric logistic regression
model to estimate propensity score. When the propensity score model
is misspecified, the weighting estimators are not consistent to the
causal effect defined on the target population $\mathcal{O}=\{x\mid\alpha<e(x)<1-\alpha\}$.
However, our estimators can still be helpful to inform treatment effects
for the population defined as $\mathcal{O}^{*}=\{x\mid\alpha<e(x'\theta^{*})<1-\alpha\}$,
where $e(x'\theta^{*})$ is the propensity score projected to the
generalized linear model family. In this case, the smooth weighting
estimators are still asymptotically linear and the bootstrap can be
used for constructing confidence intervals. See the Supplementary
Material for details. Alternatively, we can consider robust nonparametric
methods for propensity score estimation such as power series \citep{hirano2003efficient},
boosting trees, and random forest \citep{lee2010improving}.

\bibliographystyle{dcu}
\bibliography{ci}

@Article{abadie2016matching,
  Title                    = {Matching on the estimated propensity score},
  Author                   = {Abadie, Alberto and Imbens, Guido W},
  Journal                  = {Econometrica},
  Year                     = {2016},
  Pages                    = {781--807},
  Volume                   = {84},
  Publisher                = {Wiley Online Library}
}

@Article{bang2005doubly,
  Title                    = {Doubly robust estimation in missing data and causal inference models},
  Author                   = {Bang, Heejung and Robins, James M},
  Journal                  = {Biometrics},
  Year                     = {2005},
  Pages                    = {962--973},
  Volume                   = {61},
  Publisher                = {Wiley Online Library}
}

@Article{crump2009dealing,
  Title                    = {Dealing with limited overlap in estimation of average treatment effects},
  Author                   = {Crump, Richard K and Hotz, V Joseph and Imbens, Guido W and Mitnik, Oscar A},
  Journal                  = {Biometrika},
  Year                     = {2009},
  Pages                    = {187--199},
  Volume                   = {96},
  Publisher                = {Biometrika Trust}
}

@Article{dehejia1999causal,
  Title                    = {Causal effects in nonexperimental studies: Reevaluating the evaluation of training programs},
  Author                   = {Dehejia, Rajeev H and Wahba, Sadek},
  Journal                  = {J Am Stat Assoc},
  Year                     = {1999},
  Pages                    = {1053--1062},
  Volume                   = {94},
  Publisher                = {Taylor \& Francis Group}
}

@Article{hainmueller2012entropy,
  Title                    = {Entropy balancing for causal effects: A multivariate reweighting method to produce balanced samples in observational studies},
  Author                   = {Hainmueller, Jens},
  Journal                  = {Political Analysis},
  Year                     = {2012},
  Pages                    = {25--46},
  Volume                   = {20},
  Publisher                = {SPM-PMSAPSA}
}

@Article{heckman1997matching,
  Title                    = {Matching as an econometric evaluation estimator: Evidence from evaluating a job training programme},
  Author                   = {Heckman, James J and Ichimura, Hidehiko and Todd, Petra E},
  Journal                  = {The Review of Economic Studies},
  Year                     = {1997},
  Pages                    = {605--654},
  Volume                   = {64},
  Publisher                = {Oxford University Press}
}

@Article{hirano2001estimation,
  Title                    = {Estimation of causal effects using propensity score weighting: An application to data on right heart catheterization},
  Author                   = {Hirano, Keisuke and Imbens, Guido W},
  Journal                  = {Health Services and Outcomes Research Methodology},
  Year                     = {2001},
  Pages                    = {259--278},
  Volume                   = {2},
  Publisher                = {Springer}
}

@Article{hirano2003efficient,
  Title                    = {Efficient estimation of average treatment effects using the estimated propensity score},
  Author                   = {Hirano, Keisuke and Imbens, Guido W and Ridder, Geert},
  Journal                  = {Econometrica},
  Year                     = {2003},
  Pages                    = {1161--1189},
  Volume                   = {71},
  Publisher                = {Wiley Online Library}
}

@Article{hsu2013calibrating,
  Title                    = {Calibrating sensitivity analyses to observed covariates in observational studies},
  Author                   = {Hsu, Jesse Y and Small, Dylan S},
  Journal                  = {Biometrics},
  Year                     = {2013},
  Pages                    = {803--811},
  Volume                   = {69},
  Publisher                = {Wiley Online Library}
}

@Book{imbens2015causal,
  Title                    = {{Causal Inference in Statistics, Social, and Biomedical Sciences}},
  Author                   = {Imbens, Guido W and Rubin, Donald B},
  Publisher                = {Cambridge University Press},
  Year                     = {2015},
  Address                  = {Cambridge UK}
}

@Article{kang2007demystifying,
  Title                    = {Demystifying double robustness: A comparison of alternative strategies for estimating a population mean from incomplete data},
  Author                   = {Kang, Joseph DY and Schafer, Joseph L},
  Journal                  = {Statist. Sci.},
  Year                     = {2007},
  Pages                    = {523--539},
  Volume                   = {22},
  Publisher                = {JSTOR}
}

@Article{khan2010irregular,
  Title                    = {Irregular identification, support conditions, and inverse weight estimation},
  Author                   = {Khan, Shakeeb and Tamer, Elie},
  Journal                  = {Econometrica},
  Year                     = {2010},
  Pages                    = {2021--2042},
  Volume                   = {78},
  Publisher                = {Wiley Online Library}
}

@Article{lalonde1986evaluating,
  Title                    = {Evaluating the econometric evaluations of training programs with experimental data},
  Author                   = {LaLonde, Robert J},
  Journal                  = {The American economic review},
  Year                     = {1986},
  Pages                    = {604--620},
  Publisher                = {JSTOR}
}

@Article{lee2010improving,
  Title                    = {Improving propensity score weighting using machine learning},
  Author                   = {Lee, Brian K and Lessler, Justin and Stuart, Elizabeth A},
  Journal                  = {Stat Med},
  Year                     = {2010},
  Pages                    = {337--346},
  Volume                   = {29},
  Publisher                = {Wiley Online Library}
}

@Article{li2016balancing,
  Title                    = {Balancing covariates via propensity score weighting},
  Author                   = {Li, Fan and Morgan, Kari Lock and Zaslavsky, Alan M},
  Journal                  = {J Am Stat Assoc},
  Year                     = {2016},
  Pages                    = {DOI:10.1080/01621459.2016.1260466},
  Publisher                = {Taylor \& Francis}
}

@Article{lunceford2004stratification,
  Title                    = {Stratification and weighting via the propensity score in estimation of causal treatment effects: a comparative study},
  Author                   = {Lunceford, Jared K and Davidian, Marie},
  Journal                  = {Stat Med},
  Year                     = {2004},
  Pages                    = {2937--2960},
  Volume                   = {23},
  Publisher                = {Wiley Online Library}
}

@Article{rosenbaum1983central,
  Title                    = {The central role of the propensity score in observational studies for causal effects},
  Author                   = {Rosenbaum, Paul R and Rubin, Donald B},
  Journal                  = {Biometrika},
  Year                     = {1983},
  Pages                    = {41--55},
  Volume                   = {70},
  Publisher                = {Biometrika Trust}
}

@Article{rubin1974estimating,
  Title                    = {Estimating causal effects of treatments in randomized and nonrandomized studies.},
  Author                   = {Rubin, Donald B},
  Journal                  = {J Educ Psychol},
  Year                     = {1974},
  Pages                    = {688--701},
  Volume                   = {66},
  Publisher                = {American Psychological Association}
}

@Article{rubin1977assignment,
  Title                    = {Assignment to Treatment Group on the Basis of a Covariate},
  Author                   = {Rubin, Donald B},
  Journal                  = {Journal of Educational and Behavioral statistics},
  Year                     = {1977},
  Pages                    = {1--26},
  Volume                   = {2},
  Publisher                = {Sage Publications}
}

@Article{rubin1992affinely,
  Title                    = {Affinely invariant matching methods with ellipsoidal distributions},
  Author                   = {Rubin, Donald B and Thomas, Neal},
  Journal                  = {Ann. Statist.},
  Year                     = {1992},
  Pages                    = {1079--1093},
  Volume                   = {20},
  Publisher                = {JSTOR}
}

@Book{shao2012jackknife,
  Title                    = {{The Jackknife and Bootstrap}},
  Author                   = {Shao, Jun and Tu, Dongsheng},
  Publisher                = {Springer},
  Year                     = {2012},
  Address                  = {New York}
}

\section*{Supplementary Material}

Supplementary material includes proofs of Theorems 1 and 2, optimal
support for the average treatment effect on the treated, asymptotic
linearity under model misspecification, two applications, and figures.

\global\long\def\theequation{S\arabic{equation}}
 \setcounter{equation}{0} 

\global\long\def\thesection{S\arabic{section}}
 \setcounter{equation}{0} 

\global\long\def\thetable{S\arabic{table}}
 \setcounter{equation}{0} 

\global\long\def\thefigure{S\arabic{figure}}
 \setcounter{equation}{0} 

\global\long\def\theexample{S\arabic{example}}
 \setcounter{equation}{0} 

\global\long\def\thetheorem{S\arabic{theorem}}
 \setcounter{equation}{0} 

\global\long\def\thecondition{S\arabic{condition}}
 \setcounter{equation}{0} 

\global\long\def\theremark{S\arabic{remark}}
 \setcounter{equation}{0} 

\global\long\def\thestep{S\arabic{step}}
 \setcounter{equation}{0} 

\global\long\def\theassumption{S\arabic{assumption}}
 \setcounter{equation}{0} 

\global\long\def\theproof{S\arabic{proof}}
 \setcounter{equation}{0} 

\section{Proof of Theorem 1}

We write
\begin{eqnarray}
\hat{\tau}_{\epsilon} & = & \hat{\tau}_{\epsilon}(\hat{\theta})\nonumber \\
 & \cong & \hat{\tau}_{\epsilon}(\theta^{*})+E\left\{ \frac{\partial\hat{\tau}_{\epsilon}(\theta^{*})}{\partial\theta'}\right\} (\hat{\theta}-\theta^{*})\label{eq:taylor2}\\
 & \cong & \frac{1}{N}\sum_{i=1}^{N}\frac{\omega_{\epsilon}(X_{i}'\theta^{*})}{E\{\omega_{\epsilon}(X'\theta^{*})\}}\left\{ \frac{A_{i}Y_{i}}{e(X_{i}'\theta^{*})}-\frac{(1-A_{i})Y_{i}}{1-e(X_{i}'\theta^{*})}\right\} \nonumber \\
 &  & +E\left\{ \frac{\partial\hat{\tau}_{\epsilon}(\theta^{*})}{\partial\theta'}\right\} \mathcal{I}_{\theta^{*}}^{-1}S(\theta^{*})\label{eq:taylor3}\\
 & = & \frac{1}{N}\sum_{i=1}^{N}\frac{\omega_{\epsilon}(X_{i}'\theta^{*})}{E\{\omega_{\epsilon}(X'\theta^{*})\}}\left\{ \frac{A_{i}Y_{i}}{e(X_{i}'\theta^{*})}-\frac{(1-A_{i})Y_{i}}{1-e(X_{i}'\theta^{*})}\right\} \nonumber \\
 &  & +B'\frac{1}{N}\sum_{i=1}^{N}X_{i}\frac{A_{i}-e(X_{i}'\theta^{*})}{e(X_{i}'\theta^{*})\{1-e(X_{i}'\theta^{*})\}}f(X_{i}'\theta^{*}),\nonumber 
\end{eqnarray}
where $C\cong D$ means $C=D+O_{p}(N^{-1/2})$, (\ref{eq:taylor2})
follows from the Taylor expansion, (\ref{eq:taylor3}) follows from
the fact that $\hat{\theta}-\theta^{*}\cong\mathcal{I}_{\theta^{*}}^{-1}S(\theta^{*})$,
and 
\begin{equation}
B'=E\left\{ \frac{\partial\hat{\tau}_{\epsilon}(\theta^{*})}{\partial\theta'}\right\} \mathcal{I}_{\theta^{*}}^{-1}.\label{eq:B}
\end{equation}
Therefore, the asymptotic linearity of $\hat{\tau}_{\epsilon}$ follows.
Moreover, 
\begin{eqnarray*}
 &  & N^{1/2}(\hat{\tau}_{\epsilon}-\tau_{\epsilon})\\
 & \cong & N^{-1/2}\sum_{i=1}^{N}\frac{\omega_{\epsilon}(X_{i}'\theta^{*})}{E\{\omega_{\epsilon}(X_{i}'\theta^{*})\}}\left[\frac{A_{i}\{Y_{i}-\mu(A_{i},X_{i})\}}{e(X_{i}'\theta^{*})}-\frac{(1-A_{i})\{Y_{i}-\mu(A_{i},X_{i})\}}{1-e(X_{i}'\theta^{*})}\right]\\
 &  & +N^{-1/2}\sum_{i=1}^{N}\frac{\omega_{\epsilon}(X_{i}'\theta^{*})}{E\{\omega_{\epsilon}(X_{i}'\theta^{*})\}}\left(\frac{\{A_{i}-e(X_{i}'\theta^{*})\}[\mu(A_{i},X_{i})-\mu\{A_{i},e(X_{i}'\theta^{*})\}]}{e(X_{i}'\theta^{*})\{1-e(X_{i}'\theta^{*})\}}\right)\\
 &  & +N^{-1/2}\sum_{i=1}^{N}\frac{\omega_{\epsilon}(X_{i}'\theta^{*})}{E\{\omega_{\epsilon}(X_{i}'\theta^{*})\}}\left[\frac{\{A_{i}-e(X_{i}'\theta^{*})\}\mu\{A_{i},e(X_{i}'\theta^{*})\}}{e(X_{i}'\theta^{*})\{1-e(X_{i}'\theta^{*})\}}-\tau\{e(X_{i}'\theta^{*})\}\right]\\
 &  & +N^{-1/2}\sum_{i=1}^{N}\left[\frac{\omega_{\epsilon}(X_{i}'\theta^{*})}{E\{\omega_{\epsilon}(X_{i}'\theta^{*})\}}\tau\{e(X_{i}'\theta^{*})\}-\tau_{\epsilon}\right]\\
 &  & +N^{-1/2}\sum_{i=1}^{N}B'X_{i}\frac{A_{i}-e(X_{i}'\theta^{*})}{e(X_{i}'\theta^{*})\{1-e(X_{i}'\theta^{*})\}}f(X_{i}'\theta^{*}),\\
 & = & T_{0}+T_{1}+T_{2}+T_{3},
\end{eqnarray*}
where $\tau\{e(X'\theta^{*})\}=E\{Y(1)-Y(0)\mid e(X'\theta^{*})\}$,
and by grouping different terms, 
\begin{eqnarray*}
T_{0} & = & N^{-1/2}\sum_{i=1}^{N}\left[\frac{\omega_{\epsilon}(X_{i}'\theta^{*})}{E\{\omega_{\epsilon}(X_{i}'\theta^{*})\}}\tau\{e(X_{i}'\theta^{*})\}-\tau_{\epsilon}\right],
\end{eqnarray*}
\begin{eqnarray*}
T_{1} & = & N^{-1/2}\sum_{i=1}^{N}\frac{\omega_{\epsilon}(X_{i}'\theta^{*})}{E\{\omega_{\epsilon}(X_{i}'\theta^{*})\}}\left[\frac{\{A_{i}-e(X_{i}'\theta^{*})\}\mu\{A_{i},e(X_{i}'\theta^{*})\}}{e(X_{i}'\theta^{*})\{1-e(X_{i}'\theta^{*})\}}-\tau\{e(X_{i}'\theta^{*})\}\right]\\
 &  & +N^{-1/2}\sum_{i=1}^{N}B'E\{X_{i}\mid e(X_{i}'\theta^{*})\}\frac{A_{i}-e(X_{i}'\theta)}{e(X_{i}'\theta)\{1-e(X_{i}'\theta)\}}f(X_{i}'\theta^{*}),
\end{eqnarray*}
\begin{eqnarray*}
T_{2} & = & N^{-1/2}\sum_{i=1}^{N}\frac{\omega_{\epsilon}(X_{i}'\theta^{*})}{E\{\omega_{\epsilon}(X_{i}'\theta^{*})\}}\left(\frac{\{A_{i}-e(X_{i}'\theta^{*})\}[\mu(A_{i},X_{i})-\mu\{A_{i},e(X_{i}'\theta^{*})\}]}{e(X_{i}'\theta^{*})\{1-e(X_{i}'\theta^{*})\}}\right)\\
 &  & +N^{-1/2}\sum_{i=1}^{N}B'[X_{i}-E\{X_{i}\mid e(X_{i}'\theta^{*})\}]\frac{A_{i}-e(X_{i}'\theta^{*})}{e(X_{i}'\theta^{*})\{1-e(X_{i}'\theta^{*})\}}f(X_{i}'\theta^{*}),
\end{eqnarray*}
\begin{multline}
T_{3}=N^{-1/2}\sum_{i=1}^{N}\frac{\omega_{\epsilon}(X_{i}'\theta^{*})}{E\{\omega_{\epsilon}(X_{i}'\theta^{*})\}}\left[\frac{A_{i}\{Y_{i}-\mu(A_{i},X_{i})\}}{e(X_{i}'\theta^{*})}\right.\\
-\left.\frac{(1-A_{i})\{Y_{i}-\mu(A_{i},X_{i})\}}{1-e(X_{i}'\theta^{*})}\right].\label{eq:T3}
\end{multline}
Define
\[
\mathcal{F}_{0}=\left\{ X_{1}'\theta^{*},\ldots,X_{N}'\theta^{*}\right\} ,\ \mathcal{F}_{1}=\left\{ A_{1},\ldots,A_{N},X_{1}'\theta^{*},\ldots,X_{N}'\theta^{*}\right\} ,
\]
\[
\mathcal{F}_{2}=\left\{ A_{1},\ldots,A_{N},X_{1}'\theta^{*},\ldots,X_{N}'\theta^{*},X_{1},\ldots,X_{N}\right\} .
\]
By the conditioning argument, we have $E(T_{0})=0$, for $k=1,\ldots,3$,
$E(T_{k})=E\{E(T_{k}\mid\mathcal{F}_{k-1})\}=0$, and for $k=1,\ldots,3$,
\begin{eqnarray*}
\mathrm{cov}(T_{0},T_{k}) & = & \mathrm{cov}\{E(T_{0}\mid\mathcal{F}_{0}),E(T_{k}\mid\mathcal{F}_{0})\}+E\{\mathrm{cov}(T_{0},T_{k}\mid\mathcal{F}_{0})\}\\
 & = & \mathrm{cov}\{E(T_{0}\mid\mathcal{F}_{0}),0\}+E\{0\}=0,
\end{eqnarray*}
for $k=2,3$, 
\begin{eqnarray*}
\mathrm{cov}(T_{1},T_{k}) & = & \mathrm{cov}\{E(T_{1}\mid\mathcal{F}_{1}),E(T_{k}\mid\mathcal{F}_{1})\}+E\{\mathrm{cov}(T_{1},T_{k}\mid\mathcal{F}_{1})\}\\
 & = & \mathrm{cov}\{E(T_{1}\mid\mathcal{F}_{1}),0\}+E\{0\}=0,
\end{eqnarray*}
and 
\begin{eqnarray*}
\mathrm{cov}(T_{2},T_{3}) & = & \mathrm{cov}\{E(T_{2}\mid\mathcal{F}_{2}),E(T_{3}\mid\mathcal{F}_{2})\}+E\{\mathrm{cov}(T_{2},T_{3}\mid\mathcal{F}_{2})\}\\
 & = & \mathrm{cov}\{E(T_{2}\mid\mathcal{F}_{2}),0\}+E\{0\}=0.
\end{eqnarray*}
Also, we calculate the variances of $T_{i}$, for $i=0,\ldots,3$,
as follows. For $T_{0}$, 
\[
\mathrm{var}(T_{0})=E(T_{0}^{2})=
\frac{1}{E\{\omega_{\epsilon}(X'\theta^{*})\}^{2}}	\mathrm{var}\left[   \omega_\epsilon(X' \theta^*)\tau\{e(X'\theta^{*})\}  \right].
\]
For $T_{1}$, 
\begin{eqnarray*}
\mathrm{var}(T_{1}) & = & E\{\mathrm{var}(T_{1}\mid\mathcal{F}_{0})\}=E\{E(T_{1}^{2}\mid\mathcal{F}_{0})\}\\
 & = & \frac{1}{E\{\omega_{\epsilon}(X'\theta^{*})\}^{2}}E\{\omega_{\epsilon}(X'\theta^{*})^{2}\left[\left\{ \frac{1-e(X'\theta^{*})}{e(X'\theta^{*})}\right\} ^{1/2}\mu\{1,e(X'\theta^{*})\}\right.\\
 &  & +\left.\left\{ \frac{e(X'\theta^{*})}{1-e(X'\theta^{*})}\right\} ^{1/2}\mu\{0,e(X'\theta^{*})\}\right]^{2}\\
 & + & 2\frac{1}{E\{\omega_{\epsilon}(X'\theta^{*})\}}B'E\{\omega_{\epsilon}(X'\theta^{*})E\{X\mid e(X'\theta^{*})\}\\
 &  & \times\left[\frac{\mu\{1,e(X'\theta^{*})\}}{e(X'\theta^{*})}+\frac{\mu\{0,e(X'\theta^{*})\}}{1-e(X'\theta^{*})}\right]f(X'\theta^{*})\}\\
 & + & B'E\left[f(X'\theta^{*})^{2}\frac{E\{X\mid e(X'\theta^{*})\}E\{X'\mid e(X'\theta^{*})\}}{e(X'\theta^{*})\{1-e(X'\theta^{*})\}}\right]B.
\end{eqnarray*}
For $T_{2}$, 
\begin{eqnarray*}
\mathrm{var}(T_{2}) & = & E\{\mathrm{var}(T_{2}\mid\mathcal{F}_{1})\}=E\{E(T_{2}^{2}\mid\mathcal{F}_{1})\}\\
 & = & \frac{1}{E\{\omega_{\epsilon}(X'\theta^{*})\}^{2}}E\left\{ \omega_{\epsilon}(X'\theta^{*})^{2}\left[\frac{\sigma^{2}\{1,e(X'\theta^{*})\}}{e(X'\theta^{*})}+\frac{\sigma^{2}\{0,e(X'\theta^{*})\}}{1-e(X'\theta^{*})}\right]\right\} \\
 &  & +2\frac{1}{E\{\omega_{\epsilon}(X'\theta^{*})\}}B'E\left\{ \omega_{\epsilon}(X'\theta^{*})f(X'\theta^{*})\left[\frac{\mathrm{cov}\{X,\mu(1,X)\mid e(X'\theta^{*})\}}{e(X'\theta^{*})}\right.\right.\\
 &  & +\left.\left.\frac{\mathrm{cov}\{X,\mu(0,X)\mid e(X'\theta^{*})\}}{1-e(X'\theta^{*})}\right]\right\} \\
 &  & +B'E\left[f(X'\theta^{*})^{2}\frac{\mathrm{var}\{X\mid e(X'\theta^{*})\}}{e(X'\theta^{*})\{1-e(X'\theta^{*})\}}\right]B.
\end{eqnarray*}
For $T_{3},$ 
\begin{eqnarray*}
\mathrm{var}(T_{3}) & = & E\{\mathrm{var}(T_{3}\mid\mathcal{F}_{2})\}=E\{E(T_{3}^{2}\mid\mathcal{F}_{2})\}\\
 & \cong & \frac{1}{E\{\omega_{\epsilon}(X'\theta^{*})\}^{2}}E\left[\omega_{\epsilon}(X'\theta^{*})^{2}\left\{ \frac{\sigma_{1}^{2}(X)}{e(X'\theta^{*})}+\frac{\sigma_{0}^{2}(X)}{1-e(X'\theta^{*})}\right\} \right].
\end{eqnarray*}
Because
\begin{eqnarray*}
\frac{\partial\hat{\tau}_{\epsilon}(\theta^{*})}{\partial\theta'} & = & \frac{1}{N}\sum_{i=1}^{N}\frac{\partial}{\partial\theta'}\left[\frac{\omega_{\epsilon}(X_{i}'\theta^{*})}{E\{\omega_{\epsilon}(X_{i}'\theta^{*})\}}\right]\left\{ \frac{A_{i}Y_{i}}{e(X_{i}'\theta^{*})}-\frac{(1-A_{i})Y_{i}}{1-e(X_{i}'\theta^{*})}\right\} \\
 &  & -\frac{1}{N}\sum_{i=1}^{N}\frac{\omega_{\epsilon}(X_{i}'\theta^{*})}{E\{\omega_{\epsilon}(X_{i}'\theta^{*})\}}\left[\frac{A_{i}Y_{i}}{e(X_{i}'\theta^{*})^{2}}+\frac{(1-A_{i})Y_{i}}{\{1-e(X_{i}'\theta^{*})\}^{2}}\right]f(X_{i}'\theta^{*})X_{i},
\end{eqnarray*}
we have
\begin{eqnarray*}
E\left\{ \frac{\partial\hat{\tau}_{\epsilon}(\theta^{*})}{\partial\theta}\right\}  & = & E\left(\frac{\partial}{\partial\theta}\left[\frac{\omega_{\epsilon}(X'\theta^{*})}{E\{\omega_{\epsilon}(X'\theta^{*})\}}\right]\tau(X)\right)-\frac{1}{E\{\omega_{\epsilon}(X'\theta^{*})\}}E\left\{ \omega_{\epsilon}(X'\theta^{*})f(X'\theta^{*})\vphantom{\frac{E}{1-e}}\right.\\
 &  & \times\left.\left[\frac{E\{X,\mu(1,X)\mid e(X'\theta^{*})\}}{e(X'\theta^{*})}+\frac{E\{X,\mu(0,X)\mid e(X'\theta^{*})\}}{1-e(X'\theta^{*})}\right]\right\} \\
 & = & b_{1,\epsilon}-b_{2,\epsilon},
\end{eqnarray*}
where $b_{1,\epsilon}$ and $b_{2,\epsilon}$ are defined in Theorem
1. Therefore, according to (\ref{eq:B}), $B=(b_{1,\epsilon}-b_{2,\epsilon})'\mathcal{I}_{\theta}^{-1}$.
As a result, \textbf{\textcolor{black}{
\begin{eqnarray}
 &  & \mathrm{var}(T_{0})+\mathrm{var}(T_{1})+\mathrm{var}(T_{2})+\mathrm{var}(T_{3})\nonumber \\
 & = & \frac{1}{E\{\omega_{\epsilon}(X'\theta^{*})\}^{2}}	\mathrm{var}\left[   \omega_\epsilon(X' \theta^*)\tau\{e(X'\theta^{*})\}  \right]\label{eq:line1}\\
 &  & +\frac{1}{E\{\omega_{\epsilon}(X'\theta^{*})\}^{2}}E\left\{ \omega_{\epsilon}(X'\theta^{*})^{2}\left[\left\{ \frac{1-e(X'\theta^{*})}{e(X'\theta^{*})}\right\} ^{1/2}\mu\{1,e(X'\theta^{*})\}\right.\right.\nonumber \\
 &  & +\left.\left.\left\{ \frac{e(X'\theta^{*})}{1-e(X'\theta^{*})}\right\} ^{1/2}\mu\{0,e(X'\theta^{*})\}\right]^{2}\right\} \nonumber \\
 &  & +\frac{1}{E\{\omega_{\epsilon}(X'\theta^{*})\}^{2}}E\left\{ \omega_{\epsilon}(X'\theta^{*})^{2}\left[\frac{\sigma^{2}\{1,e(X'\theta^{*})\}}{e(X'\theta^{*})}+\frac{\sigma^{2}\{0,e(X'\theta^{*})\}}{1-e(X'\theta^{*})}\right]\right\} \nonumber \\
 &  & +\frac{1}{E\{\omega_{\epsilon}(X'\theta^{*})\}^{2}}E\left[\omega_{\epsilon}(X'\theta^{*})^{2}\left\{ \frac{\sigma^{2}(1,X)}{e(X'\theta^{*})}+\frac{\sigma^{2}(0,X)}{1-e(X'\theta^{*})}\right\} \right]\label{eq:line2}\\
 &  & +2\frac{1}{E\{\omega_{\epsilon}(X'\theta^{*})\}}B'E\left\{ \omega_{\epsilon}(X'\theta^{*})f(X'\theta^{*})\left[\frac{E\{X\mu(1,X)\mid e(X'\theta^{*})\}}{e(X'\theta^{*})}\right.\right.\nonumber \\
 &  & +\left.\left.\frac{E\{X\mu(0,X)\mid e(X'\theta^{*})\}}{1-e(X'\theta^{*})}\right]\right\} +B'\mathcal{I}_{\theta^{*}}B\nonumber \\
 & = & \sigma_{\epsilon}^{2}+b_{1,\epsilon}'\mathcal{I}_{\theta^{*}}^{-1}b_{1,\epsilon}-b_{2,\epsilon}'\mathcal{I}_{\theta^{*}}^{-1}b_{2,\epsilon},\nonumber 
\end{eqnarray}
}}\textcolor{black}{where $\sigma_{\epsilon}^{2}$ is defined as the
terms from (\ref{eq:line1}) to (\ref{eq:line2}), and the last equality
follows by plugging the expression of $B$, 
\begin{eqnarray*}
2B'b_{2,\epsilon}+B'\mathcal{I}_{\theta^{*}}B & = & 2b_{1,\epsilon}'\mathcal{I}_{\theta^{*}}^{-1}b_{2,\epsilon}-2b_{2,\epsilon}'\mathcal{I}_{\theta^{*}}^{-1}b_{2,\epsilon}+(b_{1,\epsilon}+b_{2,\epsilon})'\mathcal{I}_{\theta^{*}}^{-1}(b_{1,\epsilon}+b_{2,\epsilon})\\
 & = & b_{1,\epsilon}'\mathcal{I}_{\theta^{*}}^{-1}b_{1,\epsilon}-b_{2,\epsilon}'\mathcal{I}_{\theta^{*}}^{-1}b_{2,\epsilon}.
\end{eqnarray*}
Moreover, $\sigma_{\epsilon}^{2}$ can be further simplified as }\textbf{\textcolor{black}{
\begin{eqnarray}
\sigma_{\epsilon}^{2} & = & \frac{1}{E\{\omega_{\epsilon}(X'\theta^{*})\}^{2}}	\mathrm{var}\left\{   \omega_\epsilon(X' \theta^*)\tau(X)  \right\}
\nonumber \\
 &  & +\frac{1}{E\{\omega_{\epsilon}(X'\theta^{*})\}^{2}}E\left\{ \omega_{\epsilon}(X'\theta^{*})^{2}\left[\left\{ \frac{1-e(X'\theta^{*})}{e(X'\theta^{*})}\right\} ^{1/2}\mu(1,X)^{2}\right.\right.\nonumber \\
 &  & +\left.\left.\left\{ \frac{e(X'\theta^{*})}{1-e(X'\theta^{*})}\right\} ^{1/2}\mu(0,X)\right]^{2}\right\} \\
 &  & +\frac{1}{E\{\omega_{\epsilon}(X'\theta^{*})\}^{2}}E\left[\omega_{\epsilon}(X'\theta^{*})^{2}\left\{ \frac{\sigma^{2}(1,X)}{e(X'\theta^{*})}+\frac{\sigma^{2}(0,X)}{1-e(X'\theta^{*})}\right\} \right].\label{eq:sigsq}
\end{eqnarray}
}}In addition, Assumption 3 is the moment condition for the Central
Limit Theorem. Therefore, 
\[
N^{1/2}(\hat{\tau}_{\epsilon}-\tau_{\epsilon})\rightarrow\mathcal{N}\left(0,\sigma_{\epsilon}^{2}+b_{1,\epsilon}'\mathcal{I}_{\theta^{*}}^{-1}b_{1,\epsilon}-b_{2,\epsilon}'\mathcal{I}_{\theta^{*}}^{-1}b_{2,\epsilon}\right),
\]
in distribution, as $N\rightarrow\infty$. 

\section{Proof of Theorem 2}

Let $\hat{\mu}(A_{i},X_{i})$ converge to $\tilde{\mu}(A_{i},X_{i})$
as $N\rightarrow\infty$. If the model for $\mu(A_{i},X_{i})$ is
correctly specified, $\tilde{\mu}(A_{i},X_{i})=\mu(A_{i},X_{i})$.
Write 
\begin{eqnarray*}
\hat{\tau}_{\epsilon}^{\mathrm{aug}} & = & \hat{\tau}_{\epsilon}^{\mathrm{aug}}(\hat{\theta})\cong\hat{\tau}_{\epsilon}^{\mathrm{aug}}(\theta^{*})+E\left\{ \frac{\partial\hat{\tau}_{\epsilon}^{\mathrm{aug}}(\theta^{*})}{\partial\theta'}\right\} (\hat{\theta}-\theta^{*})\\
 & \cong & \frac{1}{N}\sum_{i=1}^{N}\frac{\omega_{\epsilon}(X_{i}'\theta^{*})}{E\{\omega_{\epsilon}(X'\theta^{*})\}}\hat{\tau}^{\mathrm{dr}}(X_{i})+E\left\{ \frac{\partial\hat{\tau}_{\epsilon}^{\mathrm{aug}}(\theta^{*})}{\partial\theta'}\right\} \mathcal{I}_{\theta^{*}}^{-1}S(\theta^{*})\\
 & \cong & \frac{1}{N}\sum_{i=1}^{N}\frac{\omega_{\epsilon}(X_{i}'\theta^{*})}{E\{\omega_{\epsilon}(X'\theta^{*})\}}\left[\frac{A_{i}Y_{i}}{e(X_{i})}+\left\{ 1-\frac{A_{i}}{e(X_{i})}\right\} \tilde{\mu}(1,X_{i})\right]\\
 &  & -\frac{1}{N}\sum_{i=1}^{N}\frac{\omega_{\epsilon}(X_{i}'\theta^{*})}{E\{\omega_{\epsilon}(X'\theta^{*})\}}\left[\frac{(1-A_{i})Y_{i}}{1-e(X_{i})}+\left\{ 1-\frac{1-A_{i}}{1-e(X_{i})}\right\} \tilde{\mu}(0,X_{i})\right]\\
 &  & +\tilde{B}'\frac{1}{N}\sum_{i=1}^{N}X_{i}\frac{A_{i}-e(X_{i}'\theta^{*})}{e(X_{i}'\theta^{*})\{1-e(X_{i}'\theta^{*})\}}f(X_{i}'\theta^{*}),
\end{eqnarray*}
where 
\begin{equation}
\tilde{B}'=E\left\{ \frac{\partial\hat{\tau}_{\epsilon}^{\mathrm{aug}}(\theta^{*})}{\partial\theta'}\right\} \mathcal{I}_{\theta^{*}}^{-1}.\label{eq:B-1}
\end{equation}
Therefore, the asymptotic linearity of $\hat{\tau}_{\epsilon}^{\mathrm{aug}}$
follows. Moreover,
\begin{eqnarray*}
 &  & N^{1/2}(\hat{\tau}_{\epsilon}^{\mathrm{aug}}-\tau_{\epsilon})\\
 & \cong & N^{-1/2}\sum_{i=1}^{N}\frac{\omega_{\epsilon}(X_{i}'\theta^{*})}{E\{\omega_{\epsilon}(X_{i}'\theta^{*})\}}\left[\frac{A_{i}\{Y_{i}-\mu(A_{i},X_{i})\}}{e(X_{i}'\theta^{*})}-\frac{(1-A_{i})\{Y_{i}-\mu(A_{i},X_{i})\}}{1-e(X_{i}'\theta^{*})}\right]\\
 &  & +N^{-1/2}\sum_{i=1}^{N}\left[\frac{\omega_{\epsilon}(X_{i}'\theta^{*})}{E\{\omega_{\epsilon}(X_{i}'\theta^{*})\}}\tau(X_{i})-\tau_{\epsilon}\right]\\
 &  & +N^{-1/2}\sum_{i=1}^{N}\tilde{B}'X_{i}\frac{A_{i}-e(X_{i}'\theta^{*})}{e(X_{i}'\theta^{*})\{1-e(X_{i}'\theta^{*})\}}f(X_{i}'\theta^{*}),\\
 &  & +N^{-1/2}\sum_{i=1}^{N}\frac{\omega_{\epsilon}(X_{i}'\theta^{*})}{E\{\omega_{\epsilon}(X'\theta^{*})\}}\left\{ 1-\frac{A_{i}}{e(X_{i})}\right\} \{\tilde{\mu}(1,X_{i})-\mu(1,X_{i})\}\\
 &  & +N^{-1/2}\sum_{i=1}^{N}\frac{\omega_{\epsilon}(X_{i}'\theta^{*})}{E\{\omega_{\epsilon}(X'\theta^{*})\}}\left\{ 1-\frac{1-A_{i}}{1-e(X_{i})}\right\} \{\tilde{\mu}(0,X_{i})-\mu(0,X_{i})\}\\
 & = & \tilde{T}_{3}+\tilde{T}_{0}+\tilde{T}_{1}+\tilde{T}_{2},
\end{eqnarray*}
where $\tilde{T}_{3}=T_{3}$ is defined in (\ref{eq:T3}), 
\begin{eqnarray*}
\tilde{T}_{0} & = & N^{-1/2}\sum_{i=1}^{N}\left[\frac{\omega_{\epsilon}(X_{i}'\theta^{*})}{E\{\omega_{\epsilon}(X_{i}'\theta^{*})\}}\tau(X_{i})-\tau_{\epsilon}\right],
\end{eqnarray*}
\[
\tilde{T}_{1}=N^{-1/2}\sum_{i=1}^{N}\tilde{B}'X_{i}\frac{A_{i}-e(X_{i}'\theta)}{e(X_{i}'\theta)\{1-e(X_{i}'\theta)\}}f(X_{i}'\theta^{*}),
\]
and
\begin{eqnarray*}
\tilde{T}_{2} & = & N^{-1/2}\sum_{i=1}^{N}\frac{\omega_{\epsilon}(X_{i}'\theta^{*})}{E\{\omega_{\epsilon}(X'\theta^{*})\}}\left\{ 1-\frac{A_{i}}{e(X_{i})}\right\} \{\tilde{\mu}(1,X_{i})-\mu(1,X_{i})\}\\
 &  & +N^{-1/2}\sum_{i=1}^{N}\frac{\omega_{\epsilon}(X_{i}'\theta^{*})}{E\{\omega_{\epsilon}(X'\theta^{*})\}}\left\{ 1-\frac{1-A_{i}}{1-e(X_{i})}\right\} \{\tilde{\mu}(0,X_{i})-\mu(0,X_{i})\}.
\end{eqnarray*}
By the same argument as in the proof of Theorem 1, $E(\tilde{T}_{j})=0$,
for $j=0,\ldots,3$, and $\mathrm{cov}(\tilde{T}_{j},\tilde{T}_{k})=0$
for all $j\neq k$ expect $\mathrm{cov}(\tilde{T}_{1},\tilde{T}_{2})$.
Moreover\textbf{\textcolor{black}{,
\begin{eqnarray*}
 &  & \mathrm{var}(\tilde{T}_{3})+\mathrm{var}(\tilde{T}_{0})+\mathrm{var}(\tilde{T}_{1})+\mathrm{var}(\tilde{T}_{2})+2\mathrm{cov}(\tilde{T}_{1},\tilde{T}_{2})\\
 & = & \frac{1}{E\{\omega_{\epsilon}(X'\theta^{*})\}^{2}}E\left[\omega_{\epsilon}(X'\theta^{*})^{2}\left\{ \frac{\sigma^{2}(1,X)}{e(X'\theta^{*})}+\frac{\sigma^{2}(0,X)}{1-e(X'\theta^{*})}\right\} \right]\\
 &  & + \frac{1}{E\{\omega_{\epsilon}(X'\theta^{*})\}^{2}}	\mathrm{var}\left\{   \omega_\epsilon(X' \theta^*)\tau(X)  \right\}+\tilde{B}'\mathcal{I}_{\theta^{*}}\tilde{B}\\
 &  & +\frac{1}{E\{\omega_{\epsilon}(X'\theta^{*})\}^{2}}E\left\{ \omega_{\epsilon}(X'\theta^{*})^{2}\left[\left\{ \frac{1-e(X'\theta^{*})}{e(X'\theta^{*})}\right\} ^{1/2}\{\tilde{\mu}(1,X)-\mu(1,X)\}\right.\right.
\end{eqnarray*}
\begin{eqnarray*}
 &  & -\left.\left.\left\{ \frac{e(X'\theta^{*})}{1-e(X'\theta^{*})}\right\} ^{1/2}\{\tilde{\mu}(0,X)-\mu(0,X)\}\right]^{2}\right\} \\
 &  & +\frac{1}{E\{\omega_{\epsilon}(X'\theta^{*})\}}\tilde{B}'E\left[\omega_{\epsilon}(X'\theta^{*})Xf(X'\theta^{*})\left\{ -\frac{\tilde{\mu}(1,X_{i})-\mu(1,X_{i})}{e(X_{i})}\right\} \right]\\
 &  & +\frac{1}{E\{\omega_{\epsilon}(X'\theta^{*})\}}\tilde{B}'E\left[\omega_{\epsilon}(X'\theta^{*})Xf(X'\theta^{*})\left\{ \frac{\tilde{\mu}(0,X)-\mu(0,X)}{1-e(X_{i})}\right\} \right]\\
 & = & \tilde{\sigma}_{\epsilon}^{2}+\tilde{B}'\mathcal{I}_{\theta^{*}}\tilde{B}+\tilde{B}'(C_{0}-C_{1})\\
 & = & \tilde{\sigma}_{\epsilon}^{2}+b_{1\epsilon}'\mathcal{I}_{\theta^{*}}b_{1\epsilon}+(C_{0}+C_{1})'\mathcal{I}_{\theta^{*}}(C_{0}+C_{1})+\tilde{B}'(C_{0}-C_{1})
\end{eqnarray*}
}}\textcolor{black}{where} \textcolor{black}{$\tilde{\sigma}_{\epsilon}^{2}$,
$C_{0}$ and $C_{1}$ are defined in Theorem 2. }Because
\begin{eqnarray*}
\frac{\partial\hat{\tau}_{\epsilon}^{\mathrm{aug}}(\theta^{*})}{\partial\theta} & = & \frac{1}{N}\sum_{i=1}^{N}\frac{\partial}{\partial\theta}\left[\frac{\omega_{\epsilon}(X_{i}'\theta^{*})}{E\{\omega_{\epsilon}(X_{i}'\theta^{*})\}}\right]\hat{\tau}^{\mathrm{aug}}(X_{i})\\
 &  & -\frac{1}{N}\sum_{i=1}^{N}\frac{\omega_{\epsilon}(X_{i}'\theta^{*})}{E\{\omega_{\epsilon}(X_{i}'\theta^{*})\}}X_{i}f(X'_{i}\theta^{*})\frac{A_{i}\{Y_{i}-\tilde{\mu}(A_{i},X_{i})\}}{e(X_{i}'\theta^{*})^{2}}\\
 &  & -\frac{1}{N}\sum_{i=1}^{N}\frac{\omega_{\epsilon}(X_{i}'\theta^{*})}{E\{\omega_{\epsilon}(X_{i}'\theta^{*})\}}X_{i}f(X'_{i}\theta^{*})\frac{(1-A_{i})\{Y_{i}-\tilde{\mu}(A_{i},X_{i})\}}{\{1-e(X_{i}'\theta^{*})\}^{2}},
\end{eqnarray*}
we have
\begin{eqnarray*}
E\left\{ \frac{\partial\hat{\tau}_{\epsilon}^{\mathrm{aug}}(\theta^{*})}{\partial\theta}\right\}  & = & E\left(\frac{\partial}{\partial\theta}\left[\frac{\omega_{\epsilon}(X'\theta^{*})}{E\{\omega_{\epsilon}(X'\theta^{*})\}}\right]\tau(X)\right)\\
 &  & -\frac{1}{E\{\omega_{\epsilon}(X_{i}'\theta^{*})\}}E\left\{ \omega_{\epsilon}(X_{i}'\theta^{*})Xf(X'\theta^{*})\frac{\mu(1,X)-\tilde{\mu}(1,X)}{e(X'\theta^{*})}\right\} \\
 &  & -\frac{1}{E\{\omega_{\epsilon}(X_{i}'\theta^{*})\}}E\left\{ \omega_{\epsilon}(X_{i}'\theta^{*})Xf(X'\theta^{*})\frac{\mu(0,X)-\tilde{\mu}(0,X)}{1-e(X'\theta^{*})}\right\} \\
 &  & =b_{1,\epsilon}-C_{0}-C_{1}.
\end{eqnarray*}
Therefore, 
\[
N^{1/2}(\hat{\tau}_{\epsilon}^{\mathrm{aug}}-\tau_{\epsilon})\rightarrow\mathcal{N}\left\{ 0,\tilde{\sigma}_{\epsilon}^{2}+b_{1\epsilon}'\mathcal{I}_{\theta^{*}}b_{1\epsilon}+(C_{0}+C_{1})'\mathcal{I}_{\theta^{*}}(C_{0}+C_{1})+\tilde{B}'(C_{0}-C_{1})\right\} ,
\]
in distribution, as $N\rightarrow\infty$. 

\section{Improving overlap for the treatment effect on the treated}

Define a general weighting average treatment effect,
\begin{equation}
\tau_{\omega}(\mathcal{O})=\frac{\sum_{i:X_{i}\in\mathcal{O}}\omega(X_{i})\tau(X_{i})}{\sum_{i:X_{i}\in\mathcal{O}}\omega(X_{i})}.\label{eq:general weightingE}
\end{equation}
The efficiency bound for $\tau_{\omega}(\mathcal{O})$ is 
\begin{equation}
V_{\omega}(\mathcal{O})=\frac{1}{E\{\omega(X)\mid X\in\mathcal{O}\}^{2}}E\left[\omega(X)^{2}\left\{ \frac{\sigma^{2}(1,X)}{e(X)}+\frac{\sigma^{2}(0,X)}{1-e(X)}\right\} \mid X\in\mathcal{O}\right].\label{eq:eff var}
\end{equation}

\citet{crump2009dealing} showed that the optimal set with which $\hat{\tau}_{\omega}(\mathcal{O})$
achieves the smallest asymptotic variance over all choices of $\mathcal{O}$
is 
\begin{equation}
\mathcal{O}=\left\{ x\mid\omega(x)\left\{ \frac{\sigma^{2}(1,x)}{e(x)}+\frac{\sigma^{2}(0,x)}{1-e(x)}\right\} \leq\gamma\right\} ,\label{eq:reduced set}
\end{equation}
where $\gamma$ is defined through the following equation: 
\begin{equation}
\gamma=2\frac{E\left[\omega^{2}(X)\left\{ \frac{\sigma^{2}(1,X)}{e(X)}+\frac{\sigma^{2}(0,X)}{1-e(X)}\right\} \mid\omega(X)\left\{ \frac{\sigma^{2}(1,X)}{e(X)}+\frac{\sigma^{2}(0,X)}{1-e(X)}\right\} <\gamma\right]}{E\left[\omega(X)\mid\omega(X)\left\{ \frac{\sigma^{2}(1,X)}{e(X)}+\frac{\sigma^{2}(0,X)}{1-e(X)}\right\} <\gamma\right]}.\label{eq:gamma}
\end{equation}
We identify that the weighted estimator for the average treatment
effect on the treated is (\ref{eq:general weightingE}) with $\omega(X)=e(X)$.
Assuming that $\sigma^{2}(1,X)=\sigma^{2}(0,X)=\sigma^{2}$, the optimal
set (\ref{eq:reduced set}) reduces to $\mathcal{O}=\{x\mid1-e(x)\geq\alpha\}$
with the cut-off value $\alpha=\sigma^{2}/\gamma$.

In practice, $\alpha$ can be determined by the smallest value of
$\alpha$ that satisfy the empirical estimate of equation (\ref{eq:gamma}):
\[
\frac{1}{\alpha}=2\frac{\sum_{i=1}^{N}e^{2}(X_{i})\left\{ \frac{1}{e(X_{i})}+\frac{1}{1-e(X_{i})}\right\} 1_{\{1-e(X_{i})\geq\alpha\}}}{\sum_{i=1}^{N}e(X_{i})1_{\{1-e(X_{i})\geq\alpha\}}}.
\]

The choice of $\alpha$ in $\mathcal{O}=\{1-e(X)\geq\alpha\}$ has
two opposite effects on the asymptotic variance in (\ref{eq:eff var}).
On the one hand, as $\alpha$ increases, we reduce the denominator
of the right hand side of (\ref{eq:eff var}), $E\{\omega(X)\mid X\in\mathcal{O}\}^{2}=E\{e(X)\mid X\in\mathcal{O}\}^{2},$
and therefore increase the asymptotic variance; on the other hand,
as $\alpha$ increases, we decrease the numerator of the right hand
side of (\ref{eq:eff var}),
\begin{multline*}
E\left[\omega(X)^{2}\left\{ \frac{\sigma^{2}(1,X)}{e(X)}+\frac{\sigma^{2}(0,X)}{1-e(X)}\right\} \mid X\in\mathcal{O}\right]\\
=E\left[e(X)\sigma^{2}(1,X)+\frac{e(X)^{2}\sigma^{2}(0,X)}{1-e(X)}\mid X\in\mathcal{O}\right],
\end{multline*}
and therefore decrease the asymptotic variance. The optimal value
of $\alpha$ balances the two effects. 

\section{Asymptotic linearity when propensity score model is misspecified }

Because $\hat{\theta}$ is the solution to the score equation $S(\theta)=0$,
under certain regularity conditions, $\hat{\theta}-\theta^{*}=\mathcal{J}_{\theta^{*}}^{-1}S(\theta^{*})+o_{p}(N^{-1/2})$,
where $\mathcal{J}_{\theta^{*}}=E\{\partial S(\theta^{*})/\partial\theta'\}$.
Here, when propensity score model is misspecified, $\mathcal{J}_{\theta^{*}}$
is not necessarily equal to $\mathcal{I}_{\theta^{*}}.$ 

We write
\begin{eqnarray*}
\hat{\tau}_{\epsilon} & = & \hat{\tau}_{\epsilon}(\hat{\theta})\\
 & \cong & \hat{\tau}_{\epsilon}(\theta^{*})+E\left\{ \frac{\partial\hat{\tau}_{\epsilon}(\theta^{*})}{\partial\theta'}\right\} (\hat{\theta}-\theta^{*})\\
 & \cong & \frac{1}{N}\sum_{i=1}^{N}\frac{\omega_{\epsilon}(X_{i}'\theta^{*})}{E\{\omega_{\epsilon}(X'\theta^{*})\}}\left\{ \frac{A_{i}Y_{i}}{e(X_{i}'\theta^{*})}-\frac{(1-A_{i})Y_{i}}{1-e(X_{i}'\theta^{*})}\right\} +E\left\{ \frac{\partial\hat{\tau}_{\epsilon}(\theta^{*})}{\partial\theta'}\right\} \mathcal{J}_{\theta^{*}}^{-1}S(\theta^{*})\\
 & = & \frac{1}{N}\sum_{i=1}^{N}\frac{\omega_{\epsilon}(X_{i}'\theta^{*})}{E\{\omega_{\epsilon}(X'\theta^{*})\}}\left\{ \frac{A_{i}Y_{i}}{e(X_{i}'\theta^{*})}-\frac{(1-A_{i})Y_{i}}{1-e(X_{i}'\theta^{*})}\right\} \\
 &  & +\Gamma'\frac{1}{N}\sum_{i=1}^{N}X_{i}\frac{A_{i}-e(X_{i}'\theta^{*})}{e(X_{i}'\theta^{*})\{1-e(X_{i}'\theta^{*})\}}f(X_{i}'\theta^{*}),
\end{eqnarray*}
where 
\[
\Gamma'=E\left\{ \frac{\partial\hat{\tau}_{\epsilon}(\theta^{*})}{\partial\theta'}\right\} \mathcal{J}_{\theta^{*}}^{-1}.
\]
Therefore, the asymptotic linearity of $\hat{\tau}_{\epsilon}$ follows.

Write 
\begin{eqnarray*}
\hat{\tau}_{\epsilon}^{\mathrm{aug}} & = & \hat{\tau}_{\epsilon}^{\mathrm{aug}}(\hat{\theta})\cong\hat{\tau}_{\epsilon}^{\mathrm{aug}}(\theta^{*})+E\left\{ \frac{\partial\hat{\tau}_{\epsilon}^{\mathrm{aug}}(\theta^{*})}{\partial\theta'}\right\} (\hat{\theta}-\theta^{*})\\
 & \cong & \frac{1}{N}\sum_{i=1}^{N}\frac{\omega_{\epsilon}(X_{i}'\theta^{*})}{E\{\omega_{\epsilon}(X'\theta^{*})\}}\hat{\tau}^{\mathrm{dr}}(X_{i})+E\left\{ \frac{\partial\hat{\tau}_{\epsilon}^{\mathrm{aug}}(\theta^{*})}{\partial\theta'}\right\} \mathcal{J}_{\theta^{*}}^{-1}S(\theta^{*})\\
 & \cong & \frac{1}{N}\sum_{i=1}^{N}\frac{\omega_{\epsilon}(X_{i}'\theta^{*})}{E\{\omega_{\epsilon}(X'\theta^{*})\}}\left[\frac{A_{i}Y_{i}}{e(X_{i})}+\left\{ 1-\frac{A_{i}}{e(X_{i})}\right\} \tilde{\mu}(1,X_{i})\right]\\
 &  & -\frac{1}{N}\sum_{i=1}^{N}\frac{\omega_{\epsilon}(X_{i}'\theta^{*})}{E\{\omega_{\epsilon}(X'\theta^{*})\}}\left[\frac{(1-A_{i})Y_{i}}{1-e(X_{i})}+\left\{ 1-\frac{1-A_{i}}{1-e(X_{i})}\right\} \tilde{\mu}(0,X_{i})\right]\\
 &  & +\tilde{\Gamma}'\frac{1}{N}\sum_{i=1}^{N}X_{i}\frac{A_{i}-e(X_{i}'\theta^{*})}{e(X_{i}'\theta^{*})\{1-e(X_{i}'\theta^{*})\}}f(X_{i}'\theta^{*}),
\end{eqnarray*}
where 
\[
\tilde{\Gamma}'=E\left\{ \frac{\partial\hat{\tau}_{\epsilon}^{\mathrm{aug}}(\theta^{*})}{\partial\theta'}\right\} \mathcal{J}_{\theta^{*}}^{-1}.
\]
Therefore, the asymptotic linearity of $\hat{\tau}_{\epsilon}^{\mathrm{aug}}$
follows.

The asymptotic linearity of the weighting estimators allows for the
bootstrap to construct confidence intervals.

\section{The National Health and Nutrition Examination Survey Data}

We examine a data set from the 2007\textendash 2008 U.S. National
Health and Nutrition Examination Survey to estimate the causal effect
of smoking on blood lead levels. The data set includes $3340$ subjects
consisting of $679$ smokers, denoted as $A=1$, and $2661$ nonsmokers,
denoted as $A=0$. The outcome variable $Y$ is the measured lead
level in blood, with observed range from $0.18$ ug/dl to $33.10$
ug/dl. The covariates $X$ include age, income-to-poverty level, gender,
education and race. For details of the data set, see \citet{hsu2013calibrating}. 

\begin{figure}[h]
\begin{centering}
\includegraphics{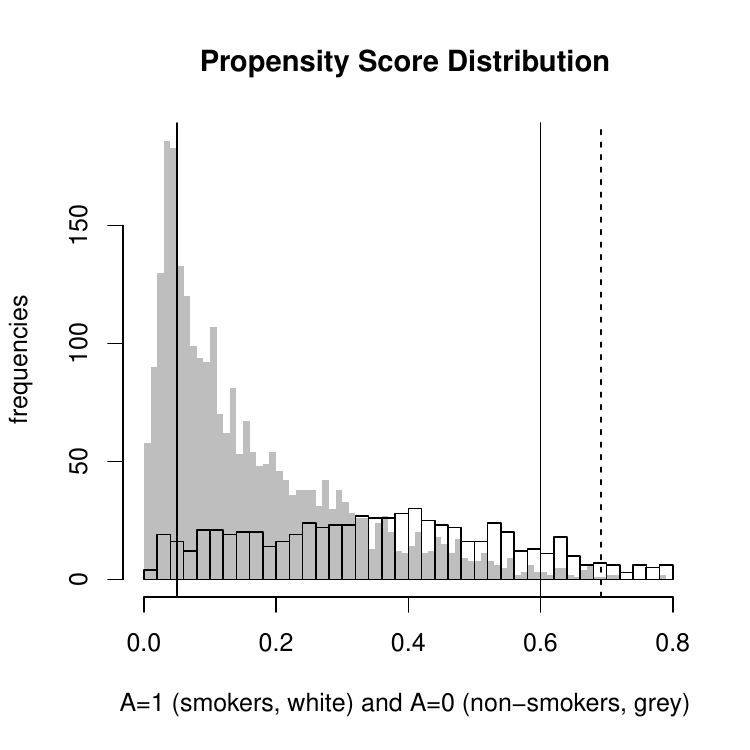}
\par\end{centering}
\caption{\label{fig:PSD}Propensity score distribution by $A$: the vertical
solid lines mark the cut-off values $0.05$ and $0.6$ for estimating
the average treatment effect, and the vertical dashed line marks the
cut-off value $0.7$ for estimating the average treatment effect on
the treated. }
\end{figure}

The propensity score is estimated by a logistic regression model with
linear predictors including all covariates. See Figure \ref{fig:PSD}
for illustration of the estimated propensity score distribution by
smokers and non-smokers. To help address lack of overlap, for the
average smoking effect, because there is little overlap for the propensity
score less than $0.05$ and greater than $0.6$, we restrict our estimand
to the target population $\mathcal{O}=\{x\mid0.05<e(x)<0.6\}$. This
results in removal of $794$ subjects ($23.8\%$ of the sample), with
$111$ smokers and $683$ non-smokers. Thus, the analysis sample includes
$2546$ subjects, with $568$ smokers and $1978$ non-smokers. For
the average smoking effect on the smokers, subjects are trimmed if
their estimated propensity score is greater than $0.7$. This results
in removal of $36$ subjects ($1.1\%$ of the sample), with $29$
smokers and $7$ non-smokers. Thus, the analysis sample includes $3304$
subjects, with $650$ smokers and $2654$ non-smokers. We consider
the weighting estimators using both indicator and smoothed inclusion
weights with $\epsilon=10^{-4}$. For the augmented weighting estimator,
we consider the outcome model to be a linear regression model adjusting
for all covariates, separately for $a=0,1$. 

Table \ref{tab:Results-2} shows the results from the four estimators
for the average smoking effect and the average smoking effect on the
smokers, based on the trimmed samples. The weighting estimators with
indicator weight function are close to the counterparts with smoothed
weight function, which have slightly smaller standard error. The augmented
weighting estimators have smaller standard error then the non-augmented
ones. From the results, on average, smoking increases the lead level
in blood at least by $0.65$ over the target population $\mathcal{O}$.
Moreover, smoking increases the lead level in blood at least by $0.79$
for smokers in the target population with $e(X)<0.7$. 

\begin{table}
\caption{\label{tab:Results-2}Results: estimate, standard error by $500$
bootstrapping, and $95\%$ confidence interval}

\centering{}%
\begin{tabular}{cccccccc}
 &  &  &  &  &  &  & \tabularnewline
 & estimate & s.e. & $95\%$ c.i. &  & estimate & s.e. & $95\%$ c.i.\tabularnewline
$\hat{\tau}(\hat{\theta})$ & 0.646 & 0.135 & (0.376, 0.916) & $\hat{\tau}_{\mathrm{ATT}}(\hat{\theta})$ & 0.796 & 0.103 & (0.591, 1.001)\tabularnewline
$\hat{\tau}^{\mathrm{aug}}(\hat{\theta})$ & 0.765 & 0.107 & (0.552, 0.978) & $\hat{\tau}_{\mathrm{ATT}}^{\mathrm{aug}}(\hat{\theta})$ & 0.793 & 0.088 & (0.616, 0.970)\tabularnewline
$\hat{\tau}_{\epsilon}(\hat{\theta})$ & 0.661 & 0.124 & (0.412, 0.909) & $\hat{\tau}_{\mathrm{ATT},\epsilon}(\hat{\theta})$ & 0.796 & 0.102 & (0.593, 0.999)\tabularnewline
$\hat{\tau}_{\epsilon}^{\mathrm{aug}}(\hat{\theta})$ & 0.763 & 0.105 & (0.554, 0.973) & $\hat{\tau}_{\mathrm{ATT},\epsilon}^{\mathrm{aug}}(\hat{\theta})$ & 0.792 & 0.088 & (0.616, 0.968)\tabularnewline
 &  &  &  &  &  &  & \tabularnewline
\end{tabular}
\end{table}

\section{The Lalonde Data}

We examine the \citet{lalonde1986evaluating} data to investigate
the treatment effect of the National Support Work Demonstration, a
labor training program, on postintervention earnings. The Lalonde
data combines the treated units from a randomized evaluation of the
National Support Work Demonstration with nonexperimental comparison
units drawn from survey datasets. The data includes $185$ treated
units and $15,992$ control units. The outcome variable $Y$ is the
postintervention earnings at year 1978. The covariates $X$ include
earnings and employment status at two preintervention years 1974 and
1975, education, age, indicators for black and Hispanic, and single
versus married. This dataset has been analyzed by many researchers;
see, e.g., \citet{dehejia1999causal,hainmueller2012entropy,imbens2015causal}. 

Because the number of control units is much larger than the number
of treated units, we first create a matched dataset where we match
each treated unit with $M=5$ control units using based on Mahalanobis
distance matching on all covariates. We then apply our methods to
the matched dataset. Following \citet{hainmueller2012entropy}, the
propensity score is estimated by a logistic regression model with
linear predictors including all covariates and their pairwise one-way
interactions, and squared terms for age and years of education. 

\begin{figure}[h]
\begin{centering}
\includegraphics{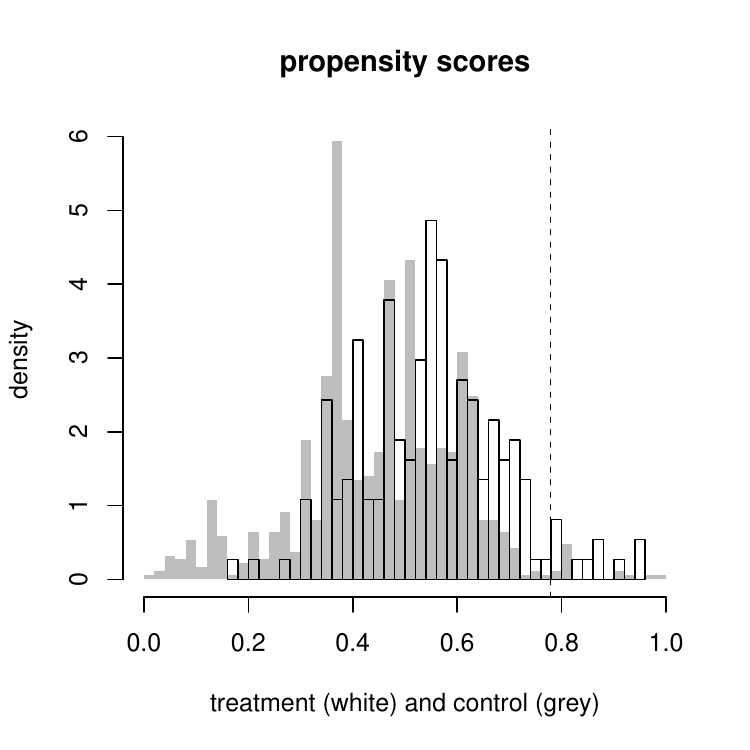}
\par\end{centering}
\caption{\label{fig:PSD-2}Propensity score distribution by $A$: the vertical
dashed line marks the cut-off value $0.75$ for estimating the average
treatment effect on the treated. }
\end{figure}

Even after matching, the overlap between the treated and the control
is not satisfactory. One implication is that for the region in the
right tail of the propensity score distribution, there is a limited
number of control units. See Figure \ref{fig:PSD-2} for illustration
of the estimated propensity score distribution by the treated and
the control. To help address lack of overlap, units are trimmed if
their estimated propensity score is greater than $0.78$, obtained
by the methods in the above section. This results in removal of $26$
subjects, with $10$ treated units and $16$ control units. We consider
the weighting estimators using both indicator and smoothed inclusion
weights with $\epsilon=10^{-4}$. For the augmented weighting estimator,
we consider the outcome model to be a linear regression model adjusting
for all covariates, and their pairwise one-way interactions, and squared
terms for age and years of education, separately for the treated and
the control.

Table \ref{tab:Results-3} shows the results from the four estimators
for the average treatment effect on the treated, based on the trimmed
samples, along with \citet{hainmueller2012entropy}'s results. The
weighting estimators with indicator weight function are close to the
counterparts with smoothed weight function, which have slightly smaller
standard error. The augmented weighting estimators do not improve
the precision of the non-augmented ones, likely because of the difficulty
in specifying a correct model for the outcome. Our point estimates
are close to the ones obtained by \citet{hainmueller2012entropy},
but the confidence intervals are much narrower, which is consistent
with the theoretical result. From the results, on average, the National
Support Work Demonstration increases the earning at least by $1305$
over the target population $\mathcal{O}$ with $e(X)<0.78$. 

\begin{table}
\caption{\label{tab:Results-3}Results: estimate, standard error by $500$
bootstrapping, and $95\%$ confidence interval}

\centering{}%
\begin{tabular}{cccc}
 &  &  & \tabularnewline
 & estimate & s.e. & $95\%$ c.i.\tabularnewline
Hain et al. & 1571 & \textendash{} & (97, 3044)\tabularnewline
$\hat{\tau}_{\mathrm{ATT}}(\hat{\theta})$ & 1506 & 404 & (733, 2321)\tabularnewline
$\hat{\tau}_{\mathrm{ATT}}^{\mathrm{aug}}(\hat{\theta})$ & 1312  & 404 & (503, 2121) \tabularnewline
$\hat{\tau}_{\mathrm{ATT},\epsilon}(\hat{\theta})$ & 1527  & 397 & (697, 2314)\tabularnewline
$\hat{\tau}_{\mathrm{ATT},\epsilon}^{\mathrm{aug}}(\hat{\theta})$ & 1305  & 400 & (505, 2014)\tabularnewline
 &  &  & \tabularnewline
\end{tabular}

Hain et al. is the results from \citet{hainmueller2012entropy}.
\end{table}

\section{Figures}

This section presents figures mentioned in the manuscript. 

\begin{figure}[h]
\begin{centering}
\includegraphics[scale=0.5]{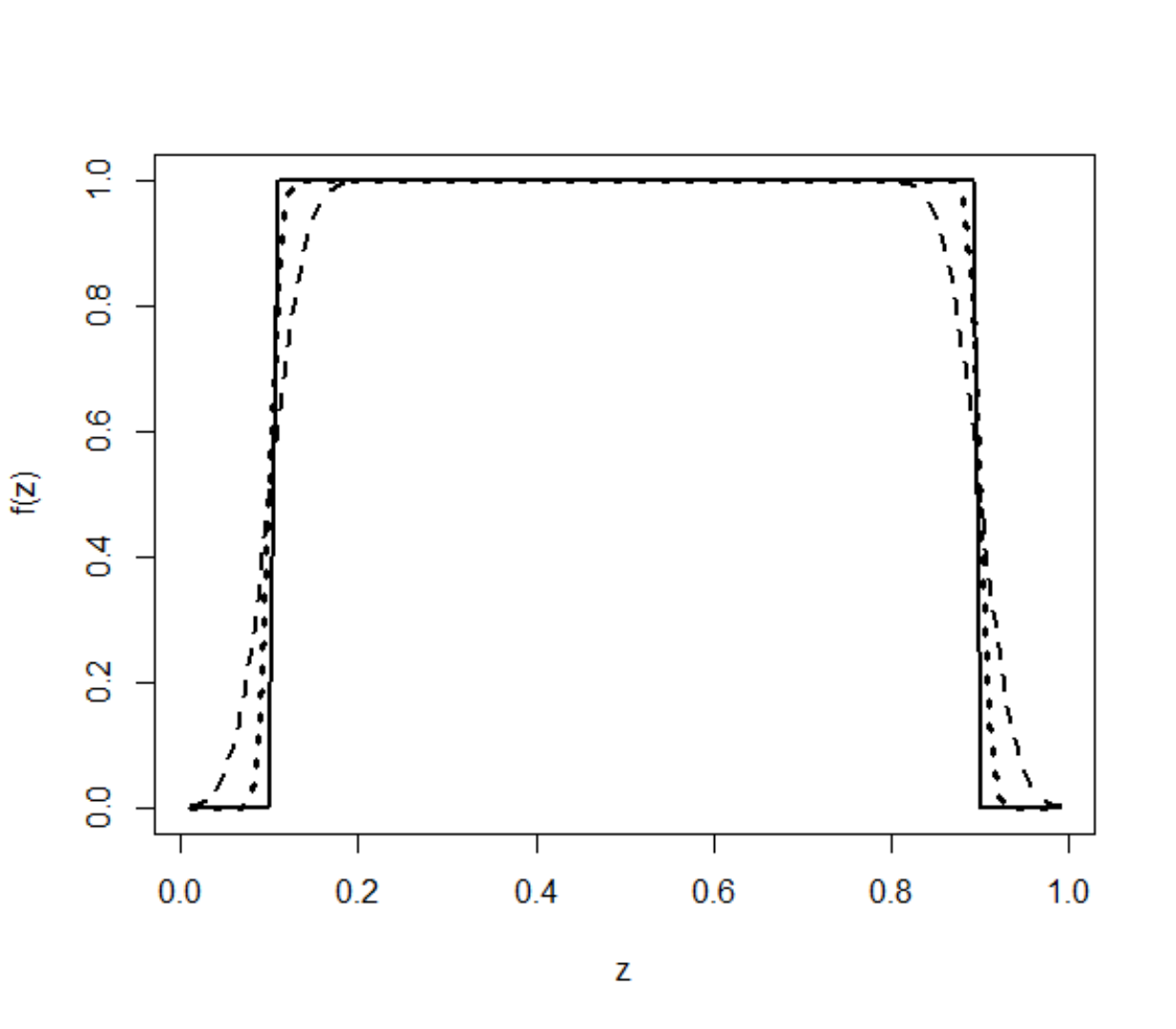}
\par\end{centering}
\caption{\label{fig:Illustration-of-weight}Illustration of weight functions:
the solid black line is the indicator weight function, the dash red
line is the smoothed weight function with $\epsilon=0.001$, and the
dot blue line is the smoothed weight function with $\epsilon=0.0001$.
As $\epsilon\rightarrow0$, the smoothed weight function converges
to the indicator function. }
\end{figure}

\begin{figure}[h]
\begin{centering}
\includegraphics[scale=0.3]{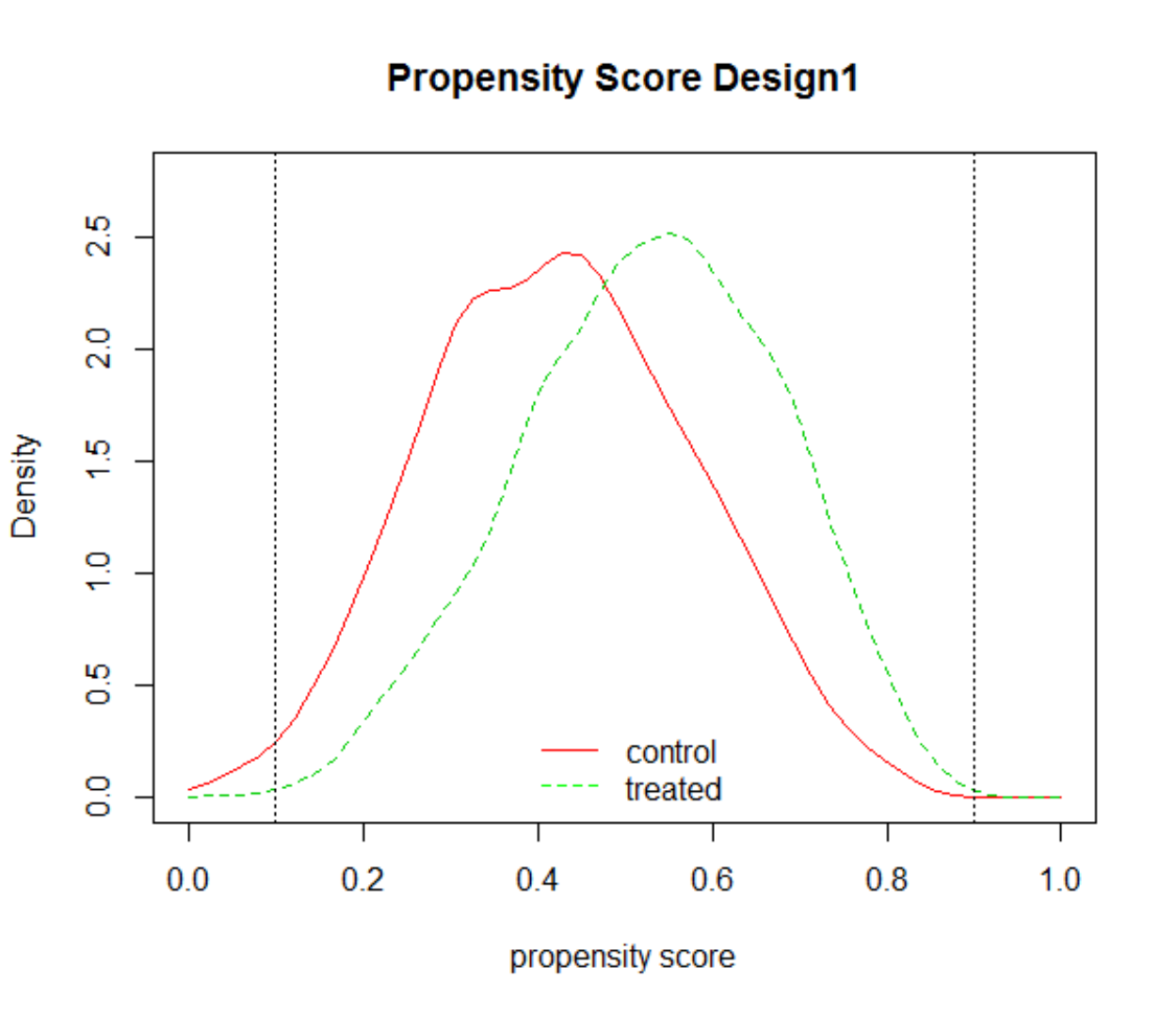}\includegraphics[scale=0.3]{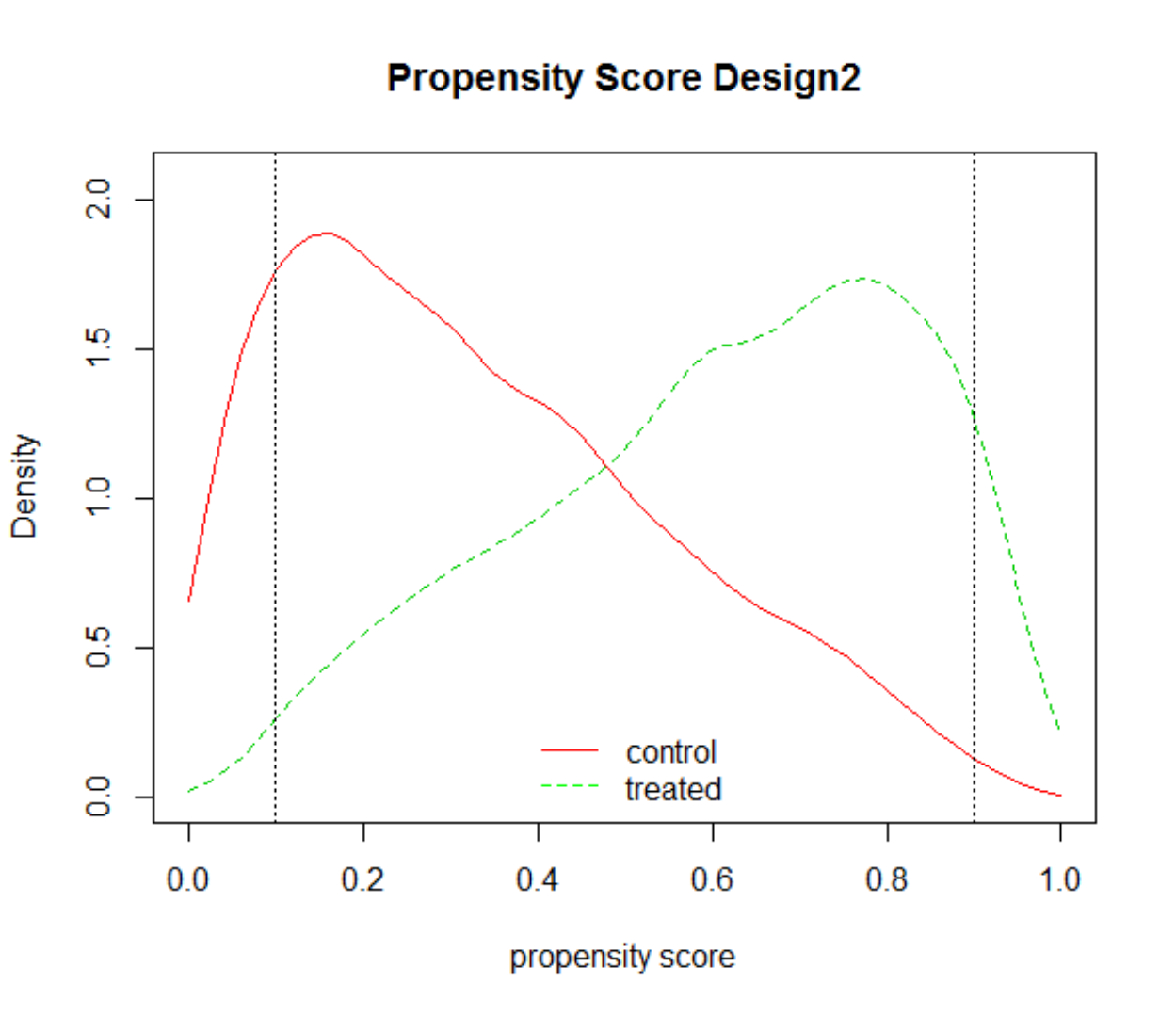}
\par\end{centering}
\begin{centering}
\includegraphics[scale=0.3]{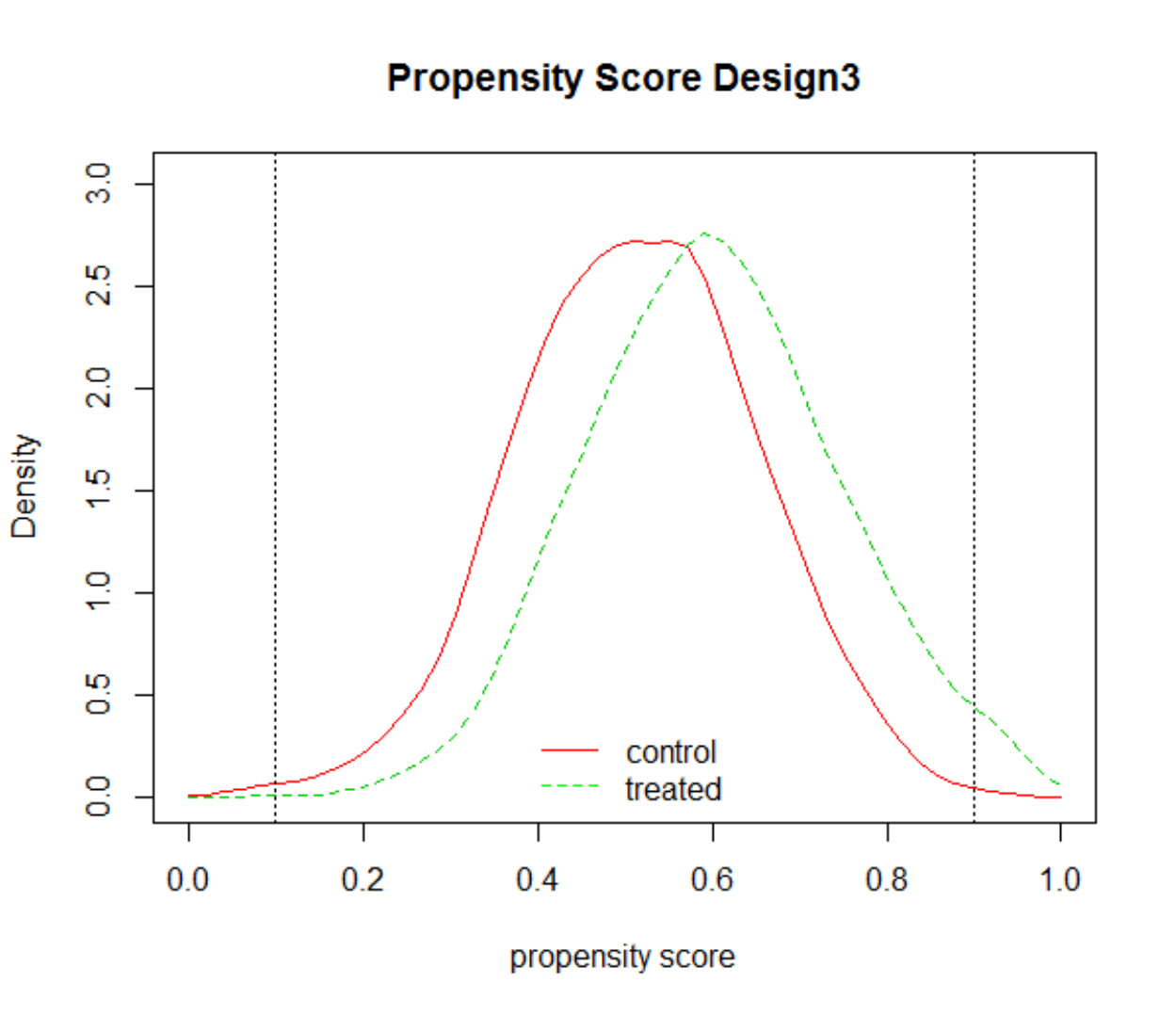}\includegraphics[scale=0.3]{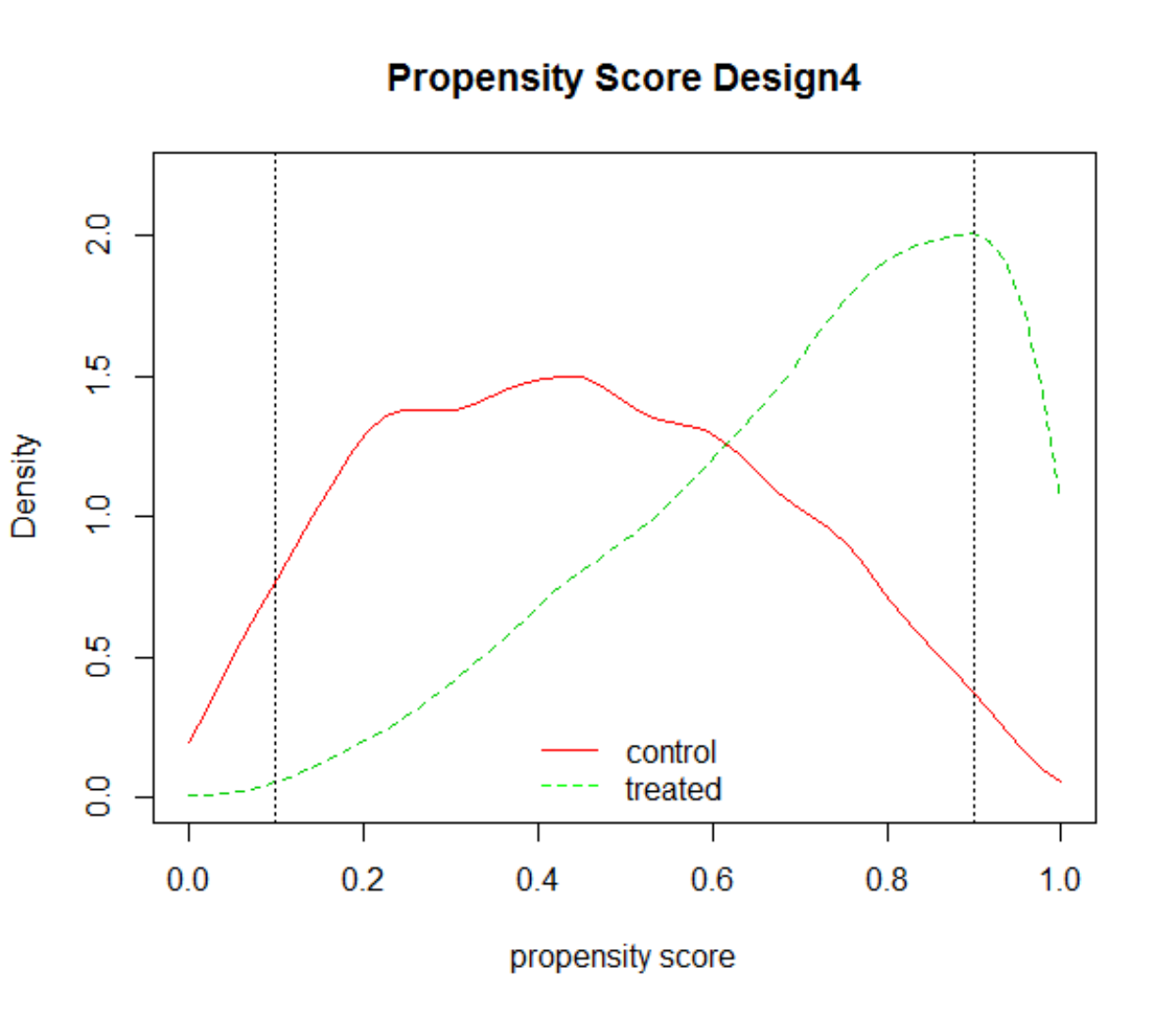}
\par\end{centering}
\caption{\label{fig:Propensity-score-distribution}Propensity score distribution
by treatment under Propensity Score Design 1\textendash 4: Design
1, weak separation and linear predictor; Design 2, strong separation
and linear predictor; Design 3, weak separation and non-linear predictor;
Design 4, strong separation and non-linear predictor.}
\end{figure}

\end{document}